\documentclass[useAMS,usenatbib]{mn2e}

\usepackage{graphicx}

\title[Secular Evolution and Bar Structure]{Secular Evolution and Structural Properties of Stellar Bars in Galaxies}
\author[Dimitri A. Gadotti]{Dimitri A. Gadotti\thanks{E-mail: dgadotti@eso.org}$^{1,2}$\\
$^1$Max-Planck-Institut f\"ur Astrophysik, Karl-Schwarzschild-Str. 1, D-85748
Garching bei M\"unchen, Germany\\
$^2$European Southern Observatory, Casilla 19001, Santiago 19, Chile}
\begin{document}

\date{Accepted 2011 April 20.  Received 2011 April 19; in original form 2010 March 6}

\pagerange{\pageref{firstpage}--\pageref{lastpage}} \pubyear{2008}

\maketitle

\label{firstpage}

\begin{abstract}
I present results from the modeling of stellar bars in nearly 300 barred galaxies in the local universe
through parametric multi-component multi-band image fitting. The surface brightness radial profile of bars is
described using a S\'ersic function, and parameters such as bar effective radius, ellipticity, boxiness,
length and mass, and bar-to-total luminosity and mass ratios, are determined,
which is unprecedented for a sample of this size.
The properties of bars in galaxies with classical bulges and pseudo-bulges are compared.
For a fixed bar-to-total mass ratio, pseudo-bulges are on average significantly less massive than classical bulges, indicating that, if pseudo-bulges are formed through bars, further processes are necessary to build a classical bulge.
I find a correlation between bar ellipticity and boxiness, and define a new parameter as the product of these two quantities. I also find correlations between this product and normalised bar
size, between the sizes of bars and bulges,
and between normalised bar size and bulge-to-total ratio. Bars with different ellipticities follow
parallel lines in the latter two correlations.
These correlations can arise if, starting off with different normalised sizes and ellipticities,
bars grow longer and stronger with dynamical age, as a result of angular momentum exchange from the inner
to the outer parts of galaxies, consistent with previous theoretical predictions. A plausible consequence is that
bar pattern speeds should become lower with bar dynamical age, and towards galaxies with more prominent
bulges.
\end{abstract}


\section{Introduction}
\label{sec:intro}

Many recent studies, from observational and theoretical viewpoints,
have established that stellar bars in disc galaxies can play an important
role in galaxy evolution \citep[see][for reviews]{SelWil93,KorKen04,Gad09a}. Theoretical work indicates
that the redistribution of
angular momentum, induced by the bar, in the galaxy interstellar medium, as well as in the stellar and dark
matter components, has a number of important consequences \citep[e.g.][]{AthMis02,Ath03,MarShlHel06,BerShlJog06}.
Gas lying beyond the bar ends is driven outwards, whereas gas lying within the bar ends is
driven to the central regions \citep[e.g.][]{Sch81,ComGer85,Ath92,FriBen93,FriBenKen94,PinStoTeu95}.
This secular evolution scenario has been partially confirmed, at least qualitatively, with observational evidence that barred galaxies show flatter
chemical abundance (O/H) radial gradients \citep[][further, \citealt{MarRoy94} find that
the stronger the bar the flatter the gradient]{ZarKenHuc94} and higher central concentrations of molecular gas
(CO -- \citealt{SakOkuIsh99}). This movement of gas to the centre might in principle help build a young
and kinematically cold stellar bulge component, i.e. a disc-like bulge \citep[see][]{Ath05b}. Indeed,
observations suggest that disc-like bulges exist and have formation processes linked to dynamical disc
instabilities, such as bars, as opposed to the old and kinematically hot classical bulges
\citep[e.g.][and references therein]{CarStideZ97,GadDos01,ErwBelGra03,Fis06,DroFis07,FisDro08,Gad09b}.

Theory also suggests how bars evolve with time. Broadly speaking, bars slow down their pattern rotation
speed, and get longer and thinner (i.e. more eccentric and stronger) during the course of their evolution, capturing stars from the disc \citep[see][]{Ath03}. Observations suggest that the strong bar in NGC 4608 has increased in mass by a factor of $\approx1.7$, through the capture of $\approx13\%$ of the disc stars \citep{Gad08}. 
In addition, more evolved bars also show more rectangular-like face-on isophotal shapes, i.e. they
are more boxy \citep[as in the $N$-body simulations of][]{AthMis02}. In detail, however, simulated bars can become abruptly shorter and thicker a few Giga-years
after their formation, due to the onset of dynamical vertical instabilities that originate box/peanut
bulges. (These seem to be simply the inner parts of bars that buckle off the disc plane and can be seen in inclined systems.) Then, about 1 Gyr later, they recover the original evolutionary trends. Furthermore, a substantial
gas component in the disc can also complicate the picture of the evolution of bar properties, halting how these
properties change, and in some cases even reversing the trends
\citep[e.g.][]{BouCom02,BouComSem05,DebMayCar06,BerShlMar07}. To date, there is no study aimed directly at
providing observational evidence on how bar properties change with time. This is partially due to
the difficulty of estimating bar dynamical ages. Although some first steps have been done in this direction
\citep[see][]{GadDes05,GadDes06,PerSanZur09}, the results are as yet inconclusive, and the methods developed require
large amounts of telescope time.

Moreover, bars are found very often in disc
galaxies, and the fraction of disc galaxies hosting bars seems to increase with time, i.e. the fraction
is lower at redshift $z\sim1$, as compared to $z\sim0$
\citep[see][and references therein, but see also \citealt{BarJogMar08} and \citealt{JogBarRix04}]{SheElmElm08}.
In addition, weaker, but also global non-axisymmetric structures, such as oval distortions in the disc, can as well
efficiently produce such redistribution of angular momentum. Therefore, the consequences of the presence
of such structures should be conspicuous, and studies on the properties of bars and their host galaxies at
$z\approx0$, as well as higher redshifts, can give direct clues on galaxy evolution.

The first and easier step in observational studies of barred galaxies is to identify bars, and there are a number
of studies on the fraction of disc galaxies with bars \citep[e.g.][]{EskFroPog00,BarJogMar08,MarJogHei09}.
Other studies have used ellipse fits to the images of barred galaxies to obtain bar properties such as
ellipticity and length \citep[e.g.][]{MarJog07,BarJogMar08,BarJabDes09}, although it has been shown that
ellipse fits can lead to an underestimation of the bar ellipticity \citep[or an overestimation of the bar axial ratio,][]{Gad08}.
In \citet{Mar95}, bar axial ratios and lengths were visually assessed, and a relation was found between bar length and the normalised diameter of the bulge. Ten years later, \citet{Erw05} measured bar lengths using ellipse fits and found that bar size scales with disc size, and confirmed previous results that bars in early-type disc galaxies are clearly larger than those in late-type galaxies \citep[see also][]{AguMenCor09}. He went further and argued that this observational evidence can be qualitatively consistent with the simulations that show bars getting larger with time, if indeed secular evolution produces early-type disc galaxies from late-type ones \citep[see also][]{FriBen95,MarFri97,GadDes05,GadDes06}. Applying ellipse fits to 2MASS images of 151 spiral galaxies, \citet{MenSheSch07} obtained measurements of bar length and axial ratio, and found a weak trend of higher ellipticities for larger bars, which is also consistent with the results from simulations.

More recently, \citet[][see also \citealt{deJ96c,LauSalBut05,LauSalBut07,Gad08}]{DurSulBut08} used 2D bulge/bar/disc decompositions of 97 Sb, Sbc and Sc galaxies, employing $i$-band images from the Sloan Digital Sky Survey (SDSS) to obtain parameters such as bar length and bar-to-total luminosity ratio. \citet{WeiJogKho09} did a similar work using $H$-band images of 143 spirals and explored also the stellar mass content in bars. About 60\% of the galaxies in both samples are barred. In this paper, I explore the results of 2D bulge/bar/disc decompositions of 291 barred galaxies, using SDSS images in the $g$ and $i$ bands, from \citet[][hereafter Paper I]{Gad09b}. This allows me to study bar properties in a level of detail which is unprecedented for a sample of this size. A thorough characterization of bars in massive galaxies in the local universe is thus put forth in Sect. \ref{sec:comp}, after a description of the data at hand in the next section. In Sect. \ref{sec:clapse}, the properties of bars in galaxies with classical and pseudo-bulges are compared. In Sect. \ref{sec:corr}, I explore correlations between bar properties in order to test the predictions from simulations on the secular growth of bars. These results are discussed in Sect. \ref{sec:disc}, while Sect. \ref{sec:conc} summarises the paper.

\section{Data}
\label{sec:data}

The reader is referred to Paper I for a detailed account of the sample selection and
image decomposition. Here I summarise the most relevant aspects of these procedures to the
present study.

In Paper I, I have performed careful and detailed image fitting of all galaxies in a sample
of 946 systems, from bulgeless to elliptical galaxies.
The sample was designed to be concomitantly suitable for structural analysis based on
image decomposition and a fair representation of the galaxy population in the local universe. It was
drawn from all objects spectroscopically classified as galaxies in the SDSS Data Release Two (DR2)
at redshifts $0.02\leq z\leq0.07$, and with stellar masses larger than $10^{10}~{\rm M}_\odot$.
This parent sample is thus a volume-limited sample of {\em massive} galaxies, i.e. a sample which includes
all galaxies more massive than $10^{10}~{\rm M}_\odot$ in the volume defined by the redshift
cuts and the DR2 footprint. In order to produce reliable decompositions, and avoid dust and
projection effects, I have applied another important selection criterion to produce the final sample:
it contains only galaxies close
to face-on, i.e. with an axial ratio $b/a\geq0.9$, where $a$ and $b$ are, respectively, the
semi-major and semi-minor axes of the galaxy at the 25 $g$-band mag arcsec$^{-2}$ isophote.
This criterion also eases the identification of bars, which are difficult to see in very inclined projections.
I have found that the final sample is
representative of the local population of massive galaxies. This was done by comparing the distributions
of several main galaxy properties, such as absolute magnitude, D$_n$(4000)\footnote{The D$_n$(4000) index is based on the 4000\AA\, discontinuity
seen in optical spectra of galaxies, and it is not sensitive
to dust attenuation effects. Young stellar populations have
low values of D$_n$(4000), as compared to old stellar populations.} and concentration,
in the volume-limited and final samples, and verifying that these distributions are similar.

Two-dimensional fits were performed using the {\sc budda} code \citep{deSGaddos04,Gad08} and SDSS images in the $g$, $r$ and $i$ bands, including up to three components in the models, namely bulge, disc
and bar.\footnote{Results are available at
http://www.mpa-garching.mpg.de/$\sim$dimitri/buddaonsdss/buddaonsdss.html.}
The presence of these components was assessed by individual inspection of images,
surface brightness radial profiles and isophotal maps.
Given the typical redshift of the galaxies in the sample, $z\approx0.05$, and the typical FWHM
of the PSF in SDSS images, ${\rm FWHM}\approx1.5$ arcsec, the typical physical spatial
resolution in these images is thus 1.5 kpc. This means that I
likely missed most bars with semi-major axis shorter than $L_{\rm bar}\approx2-3$ kpc, typically seen in very late-type
spirals (later than Sc -- \citealt{ElmElm85}). The whole of these faint bars is typically within $2-4$ seeing
elements, and thus they do not imprint clear signatures in either the isophotal contours or the intensity profile.
The results presented here thus concern the typical, bonafide bars seen in early-type spirals and
lenticulars. Note that, in contrast, bulges with effective radius
$r_e$ of the order of one PSF HWHM can still be identified in the
intensity profile, since they usually contain a much larger fraction of the galaxy light than these short and faint
bars. The final sample includes 291 barred galaxies, which are the subject of the present paper.
Classical bulges and pseudo-bulges are separated using the \citet{Kor77} relation, where pseudo-bulges can be identified
as outliers in an objective fashion. It is worth noting that pseudo-bulges here refer to disc-like bulges, not box/peanut bulges, since the sample contains only face-on galaxies, and bars are included in the fitted models (see Paper I).

Bulge, disc and bar are described as concentric ellipses, which can have different position angles and ellipticities. They also follow a surface brightness radial profile: in the case of the disc, this is an exponential \citep{Fre70} profile. Bulge and bar follow a \citet{Ser68} profile, in which the S\'ersic index is a free parameter that controls the shape of the profile. A large S\'ersic index corresponds to a highly centrally concentrated profile with important wings, whereas a small S\'ersic index corresponds to a flatter profile with a fast declining outer part. It should be noted that, although both bulge and bar have surface brightness profiles following the same parametric function, there is little space for degenerate solutions, since bars are generally more extended and eccentric than bulges, and have lower S\'ersic indices.

Since I have done multi-band decompositions, I was able to estimate the $g-i$ integrated
colour of each component separately. Using the relation between $g-i$ and the stellar mass-to-light
ratio in the $i$-band from \citet{KauHecBud07}, I have all parameters necessary to accurately calculate
the stellar masses of all components, including the bars, of the galaxies in the sample.

The structural parameters obtained in Paper I used in this work are thus:

\begin{itemize}
\item disc scale length $h$;
\item bulge effective radius $r_e$ (i.e. the radius that contains half of the light coming from the bulge);
\item bulge-to-total luminosity ratio $B/T$;
\item bar effective radius $r_{e,bar}$ (i.e. the radius that contains half of the light coming from the bar);
\item bar-to-total luminosity ratio $Bar/T$ (lum.);
\item bar-to-total mass ratio $Bar/T$ (mass);
\item bar S\'ersic index $n_{\rm Bar}$;
\item bar ellipticity $\epsilon$, defined as $1-b/a$, where $a$ and $b$ are, respectively, the
bar semi-major and semi-minor axes;
\item bar boxiness $c$, defined through the equation of the generalised ellipse used to fit the bar
\citep[see][]{AthMorWoz90}:
\begin{equation}
\left(\frac{|x|}{a}\right)^c+\left(\frac{|y|}{b}\right)^c=1,
\end{equation}
\noindent where, again, $a$ and $b$ are, respectively, the bar semi-major and semi-minor axes, and
$x$ and $y$ are the pixel coordinates of the ellipse points;
\item bar length (semi-major axis) $L_{\rm bar}$.
\end{itemize}

\noindent It is worth noting that the bar length as determined through the image decompositions is usually larger than the radius of the peak in ellipticity inside the bar, a parameter commonly used to define bar size, estimated through ellipse fitting. However, $L_{\rm bar}$ agrees well with two other parameters used to define bar size, namely the first minimum in ellipticity outside the ellipticity peak in the bar, and the point where the position angle of the fitted ellipses differ by more than 10$^\circ$ from the position angle of the bar \citep[see][fig. 4]{Gad08}, which are known to provide more accurate estimates \citep[see discussion in][]{Erw05}.

Two other parameters used in this paper were obtained from the SDSS database:

\begin{itemize}
\item $r_{24}$: the radius of the galaxy 24 $r$-band mag arcsec$^{-2}$ isophote;
\item R90: the radius containing 90 per cent of the total galaxy light.
\end{itemize}

All these parameters refer to measurements using the $i$-band image, unless otherwise noted. In the remaining of the paper, I will explore these physical properties firstly to provide a comprehensive description of the structure of stellar bars in massive galaxies in the local universe, and, secondly, to verify if one can test the theoretical predictions described in the previous section about how such bars evolve in time.

\section{Results}
\label{sec:res}

\subsection{The structural properties of bars}
\label{sec:comp}

\begin{figure*}
   \centering
   \includegraphics[keepaspectratio=true,width=5cm,clip=true]{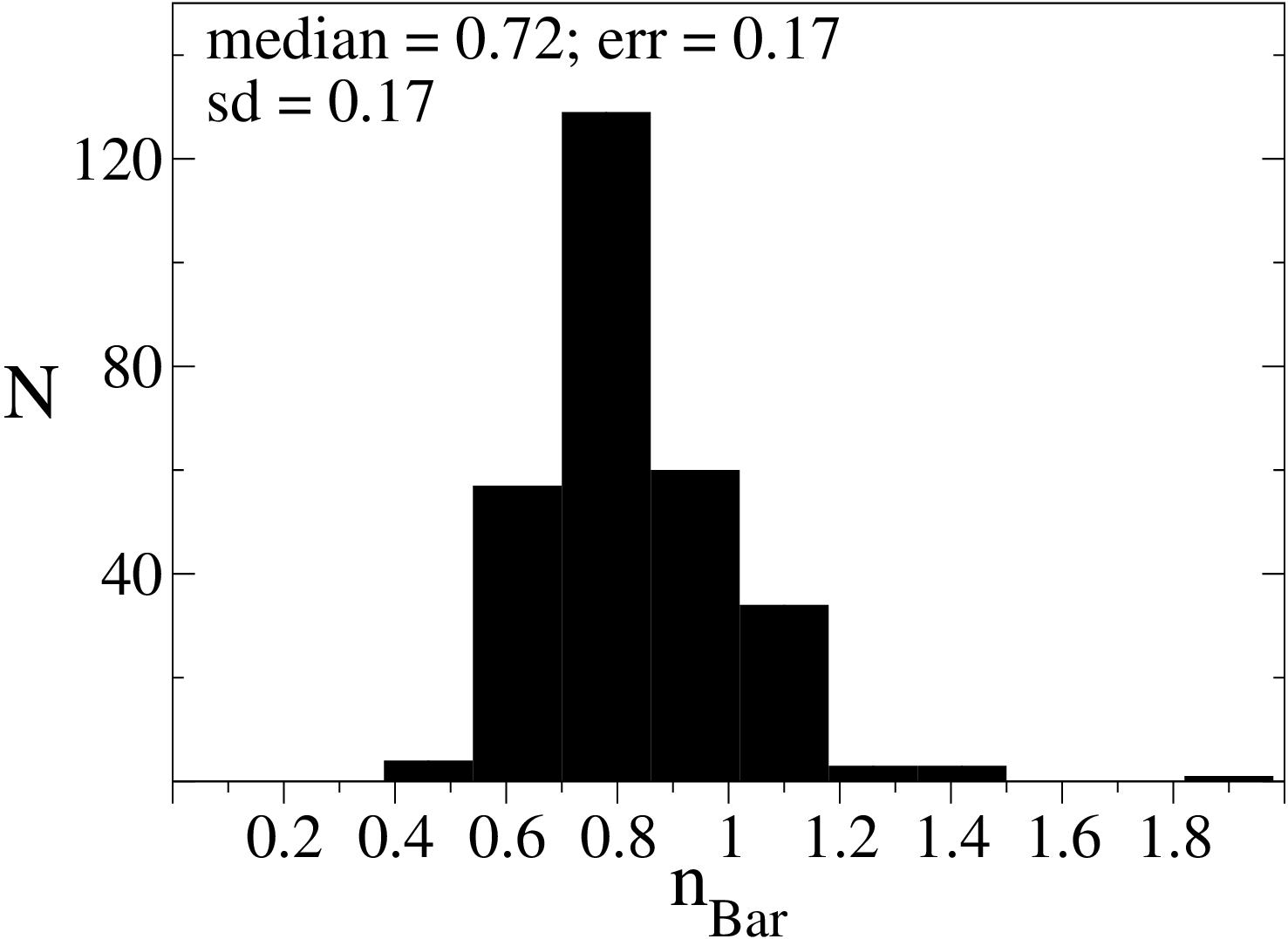}
   \includegraphics[keepaspectratio=true,width=5cm,clip=true]{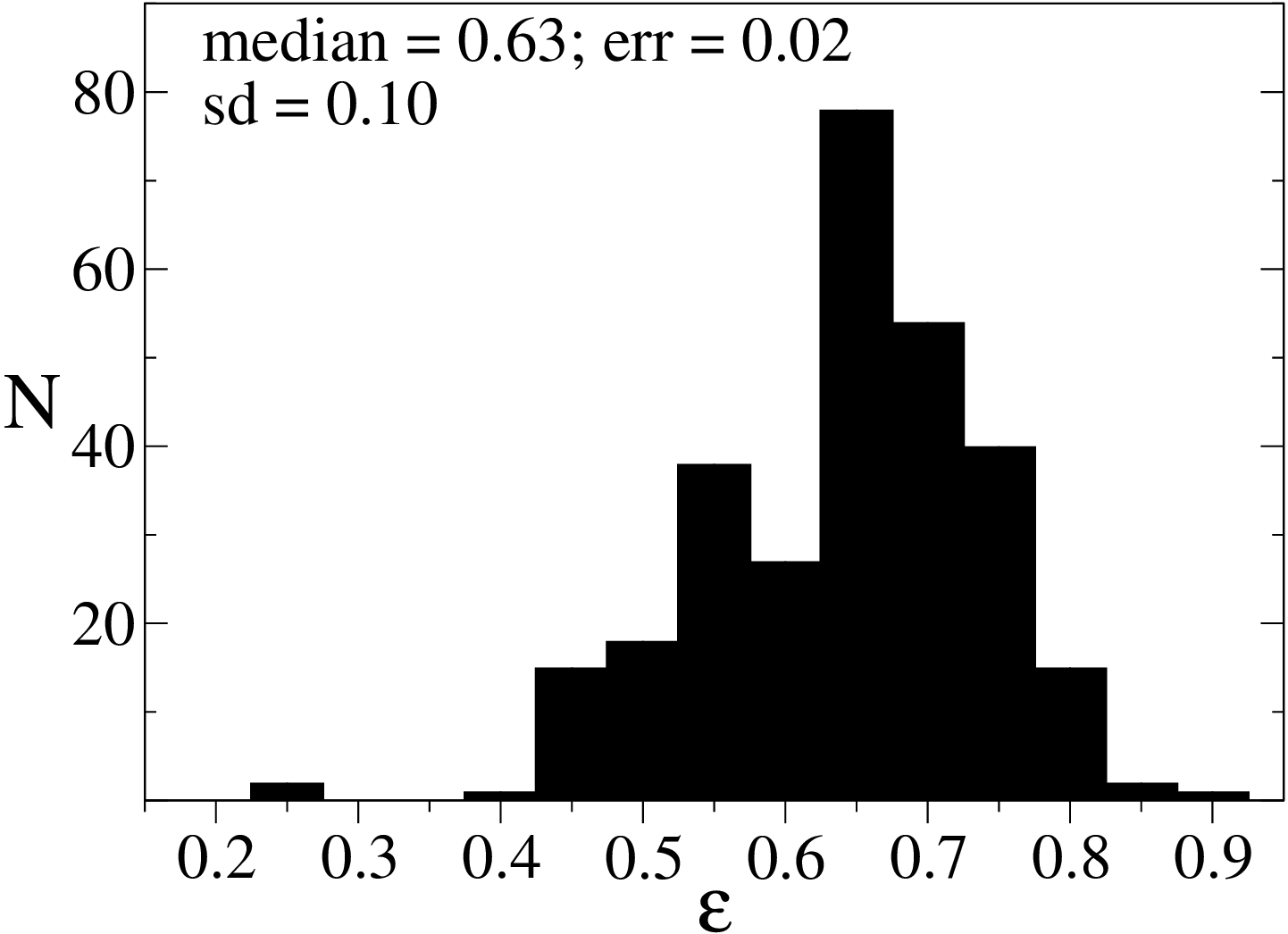}
   \includegraphics[keepaspectratio=true,width=5cm,clip=true]{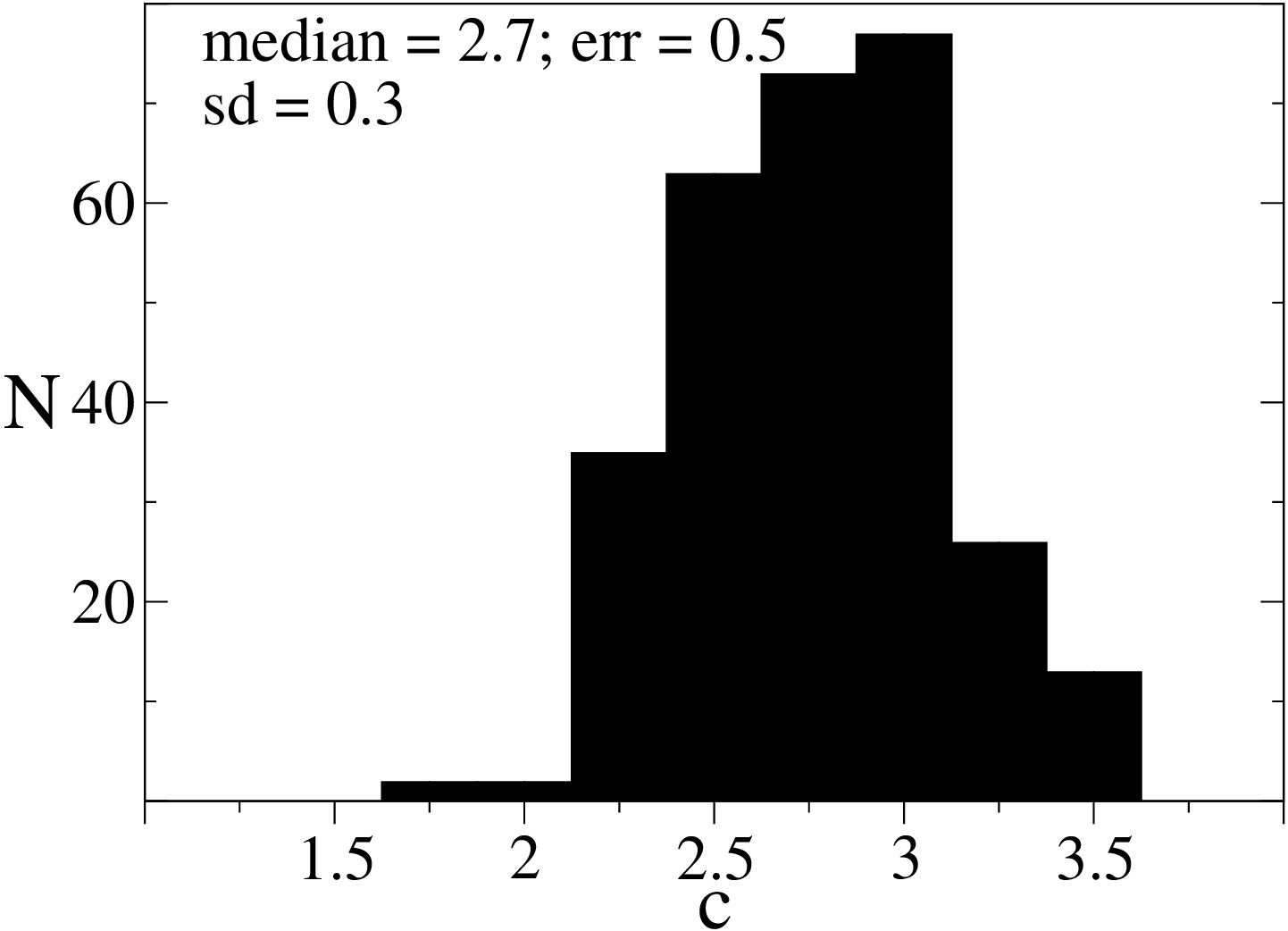}
   \includegraphics[keepaspectratio=true,width=5cm,clip=true]{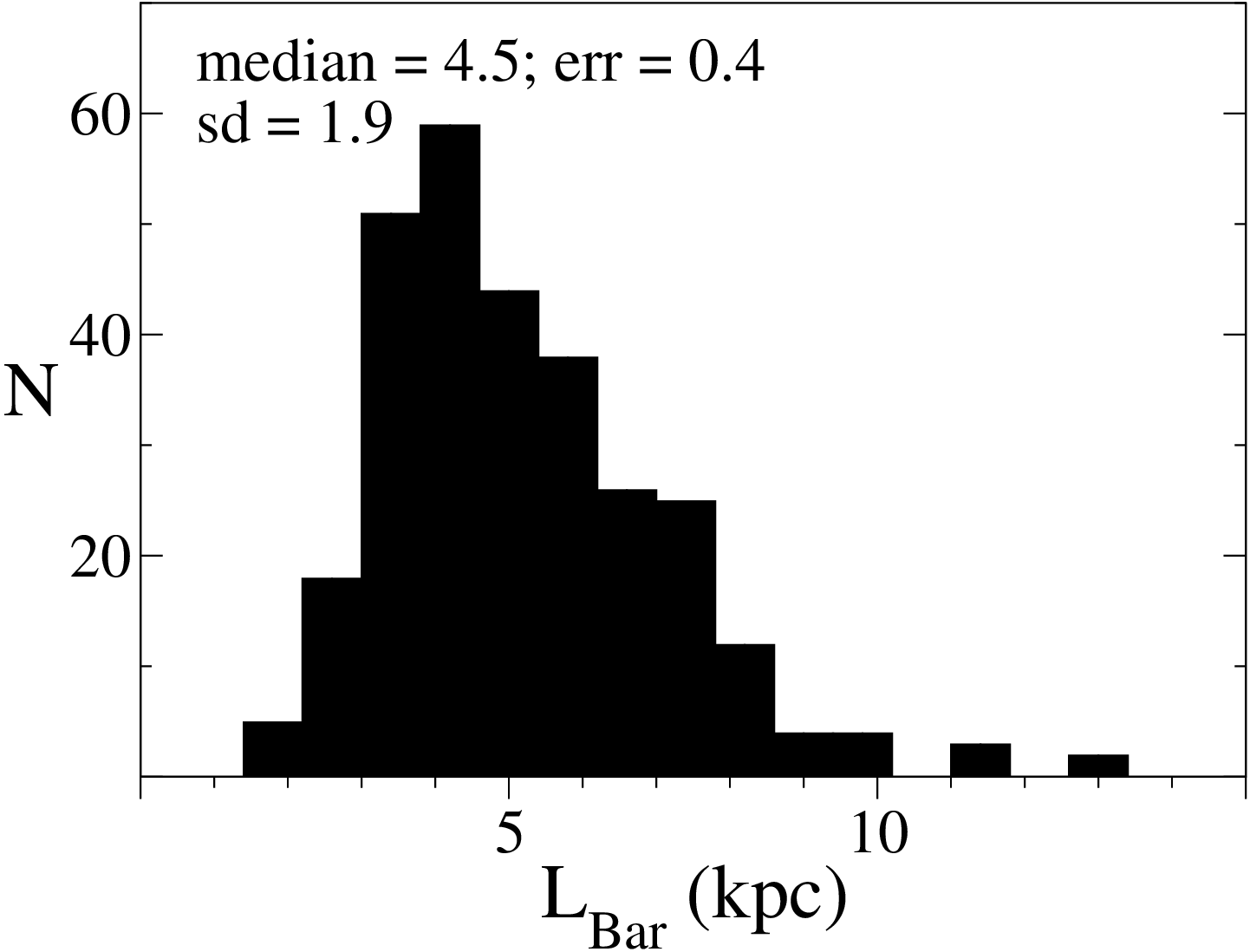}
   \includegraphics[keepaspectratio=true,width=5cm,clip=true]{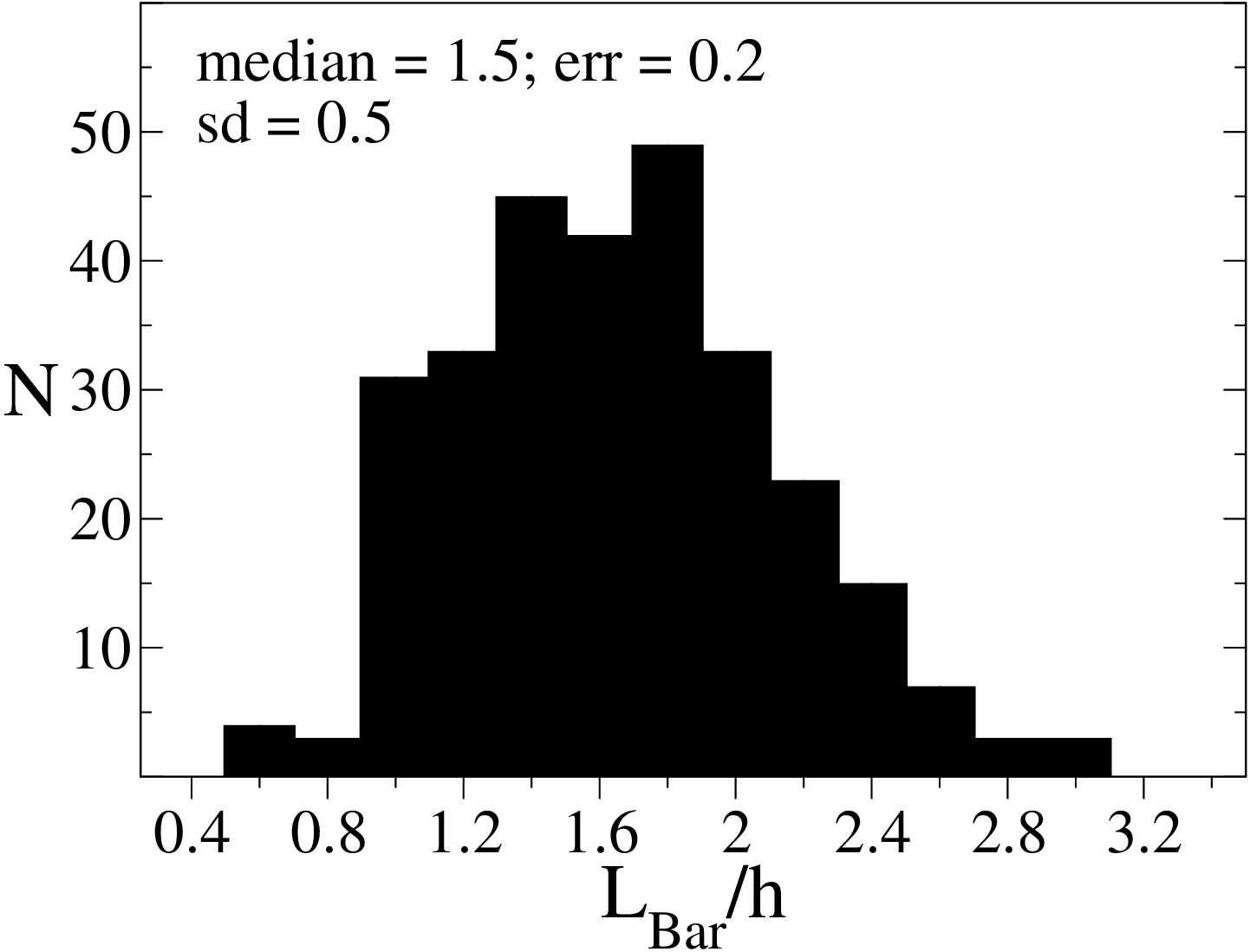}
   \includegraphics[keepaspectratio=true,width=5cm,clip=true]{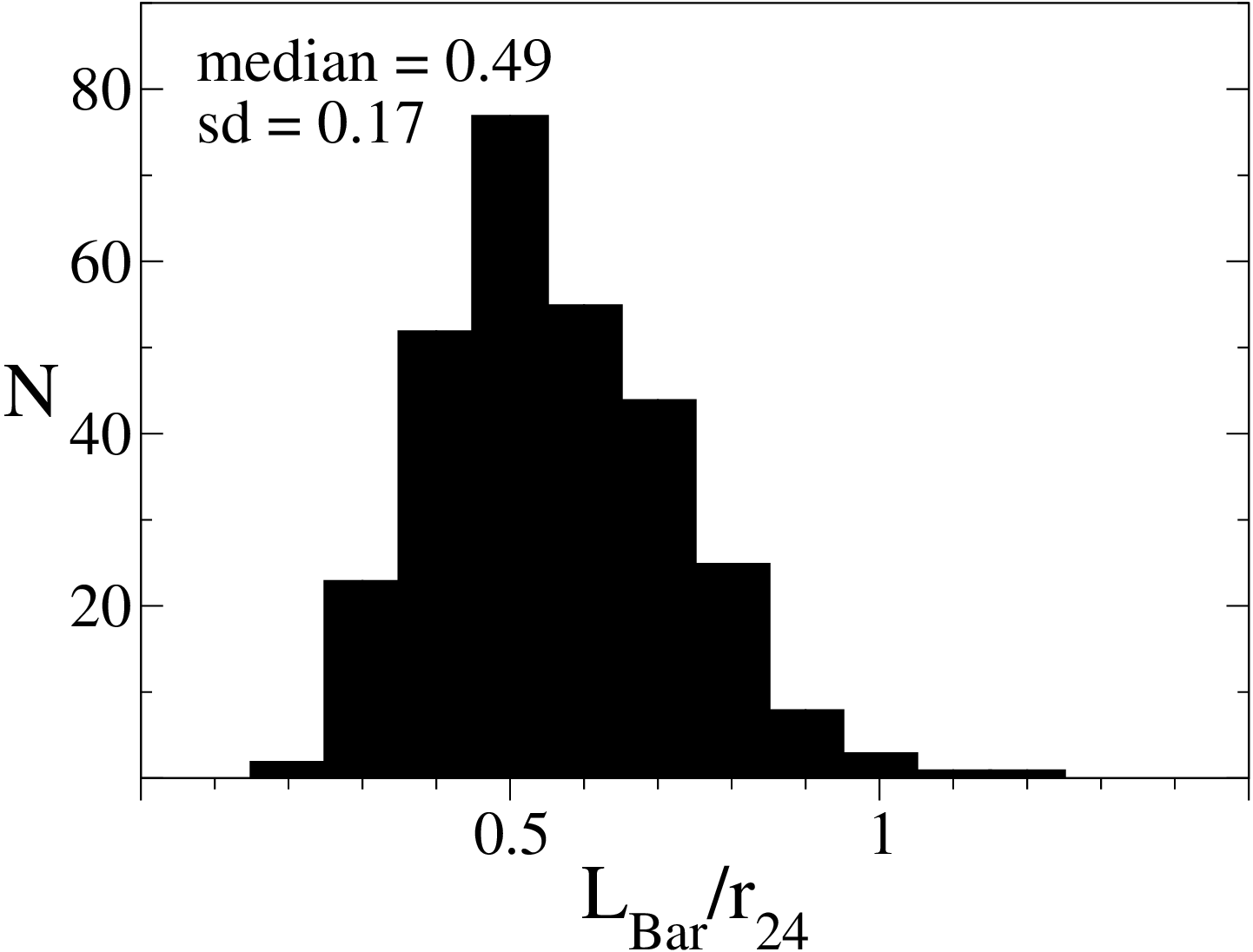}
   \includegraphics[keepaspectratio=true,width=5cm,clip=true]{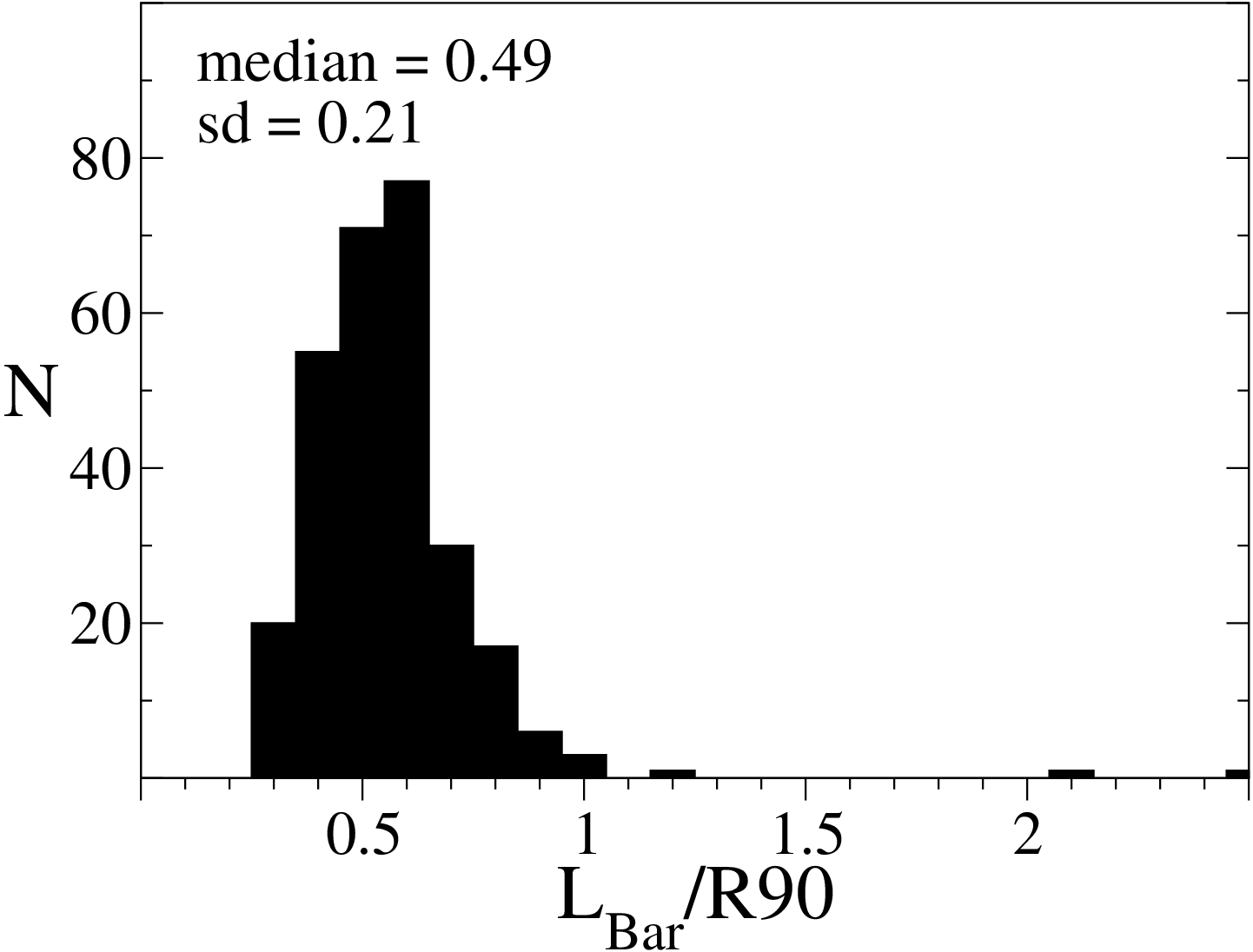}
   \includegraphics[keepaspectratio=true,width=5cm,clip=true]{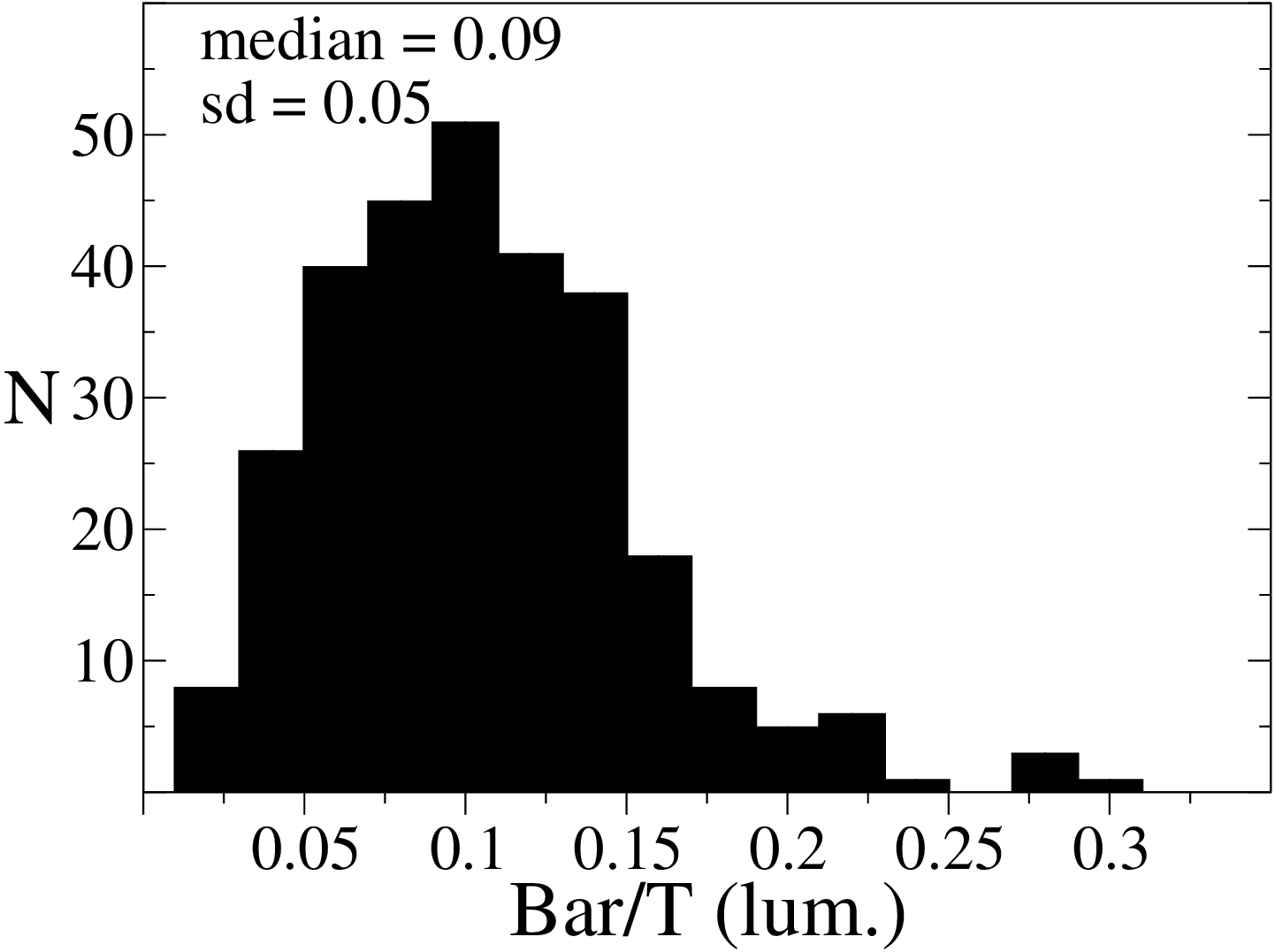}
   \includegraphics[keepaspectratio=true,width=5cm,clip=true]{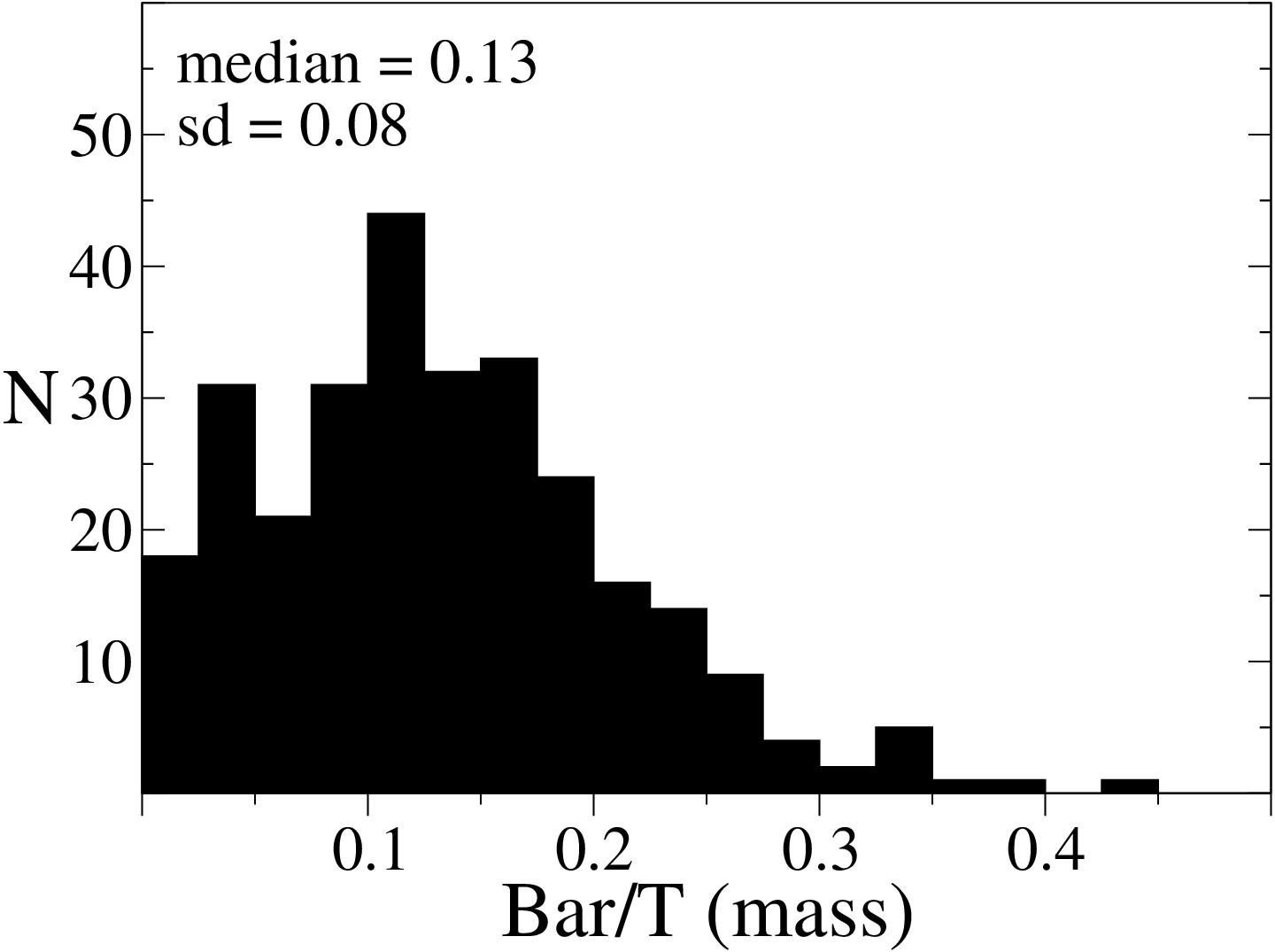}
   \caption{Distributions of several bar properties. From top to bottom and left to right:
S\'ersic index, ellipticity, boxiness, length (semi-major axis), length normalised by disc
scale-length, length normalised by the radius of the 24 $r$-band mag arcsec$^{-2}$ isophote,
length normalised by the radius containing 90 per cent of the total galaxy light,
bar-to-total luminosity ratio, and bar-to-total mass ratio. Marked at each panel are the median
and standard deviation values of the corresponding distribution, as well as the mean 1$\sigma$
error of a single measurement, when available. Bin sizes are $\approx1-2\sigma$.}
   \label{fig:sprops}
\end{figure*}

As discussed at the Introduction, large-scale stellar bars play a fundamental role in galaxy evolution. It is thus clearly very important to put forth a description of their structural properties as detailed as possible. With the data described above, I do this in Fig. \ref{fig:sprops}. This gives us a thorough portrayal of bars in the local universe, which is useful in many ways. For instance, if one wants to put bars ad hoc in a theoretical framework to describe barred galaxies today, it is most likely that one would want such bars to be consistent with those studied here. Conversely, theoretical studies aiming at the evolution of bars in time should be able to explain the existence of bars today with the properties depicted in Fig. \ref{fig:sprops}. In addition, a comparison between the properties of bars at different redshifts gives us directly a way to see how bars evolve in time.

Some features in Fig. \ref{fig:sprops} are worth mentioning. First, the typical value of the bar S\'ersic index ($n_{\rm Bar}\approx0.7$) depicts a profile which is about half way between a Gaussian (S\'ersic index equals 0.5) and an exponential (S\'ersic index equals 1). The distribution of bar ellipticities peaks at $\approx0.6$, which is about 20\% higher than the peaks found in studies based on ellipse fits \citep[e.g.][]{MenSheSch07,MarJog07,BarJogMar08,MarJogHei09}. This difference is due to the fact that the isophotes enclosing bars are slightly rounder than the bar itself, due to the contribution of the axisymmetric light distribution from bulge and disc, and thus the ellipse fits to those isophotes will also be rounder than the bar. In 2D fits, however, the light from bulge and disc is taken into account and thus a better estimate of the bar ellipticity can be obtained in this way. This effect was clearly demonstrated in \citet{Gad08}. In fact, it was shown that ellipse fits indicate bar ellipticities which are, on average, 20\% lower than those measured using 2D fits, employing the same galaxy images. Furthermore, \citet{MarJogHei09} found that their ellipticity estimates are higher on average in bulgeless galaxies. Also, most bars are not well fitted with a pure ellipse, i.e. a generalised ellipse with $c=2$, since bar boxiness peaks at $c=3$. This is important to keep in mind when one does 2D fitting of images of barred galaxies.

The distribution of bar length peaks at values which are similar to those found by \citet{MarJog07}, \citet{DurSulBut08} and \citet{AguMenCor09}. However, the distribution presented by \citet{BarJogMar08} peaks at a value which is about twice as low. This difference can likely be explained by differences in the samples and in the way bar length is defined. The sample used in  \citet{BarJogMar08} is at a redshift range of $0.01<z<0.03$, while the sample studied here is at $0.02<z<0.07$. Since they also used SDSS images, this means they have on average better spatial resolution to resolve shorter bars. In addition, they used the position of the peak in the ellipticity profile in the bar as a measure of bar length, which is known to usually result in lower values than other definitions of bar length \citep[see e.g. discussion in][]{AthMis02,Erw05}. It is also interesting to note that no bar extends to a radius larger than $\approx3$ times the disc scale length, and that almost all bars are shorter than $r_{24}$ and R90. The distribution of the bar-to-total luminosity fraction shows a peak at $\approx0.1$. The corresponding peak for the {\em mass} fraction is at a higher value, as expected, since bars are usually mostly made up of old stars. As one can see, bars can contain up to about 40\% of the total galaxy stellar mass.

\subsection{Bars in galaxies with classical and pseudo-bulges}
\label{sec:clapse}

\begin{figure*}
   \centering
   \includegraphics[keepaspectratio=true,width=7cm,clip=true]{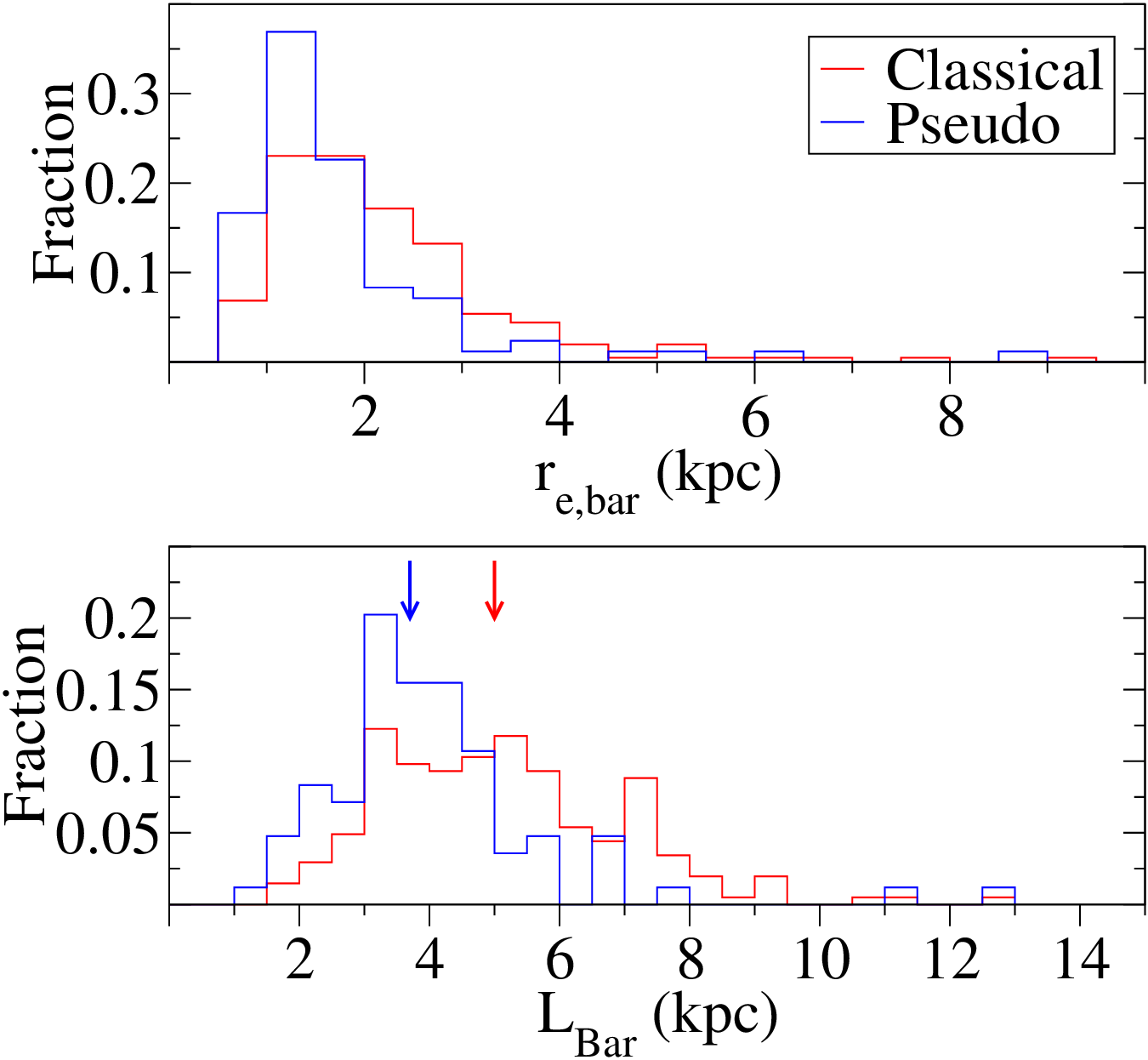}
   \hskip 0.5cm\includegraphics[keepaspectratio=true,width=10cm,clip=true]{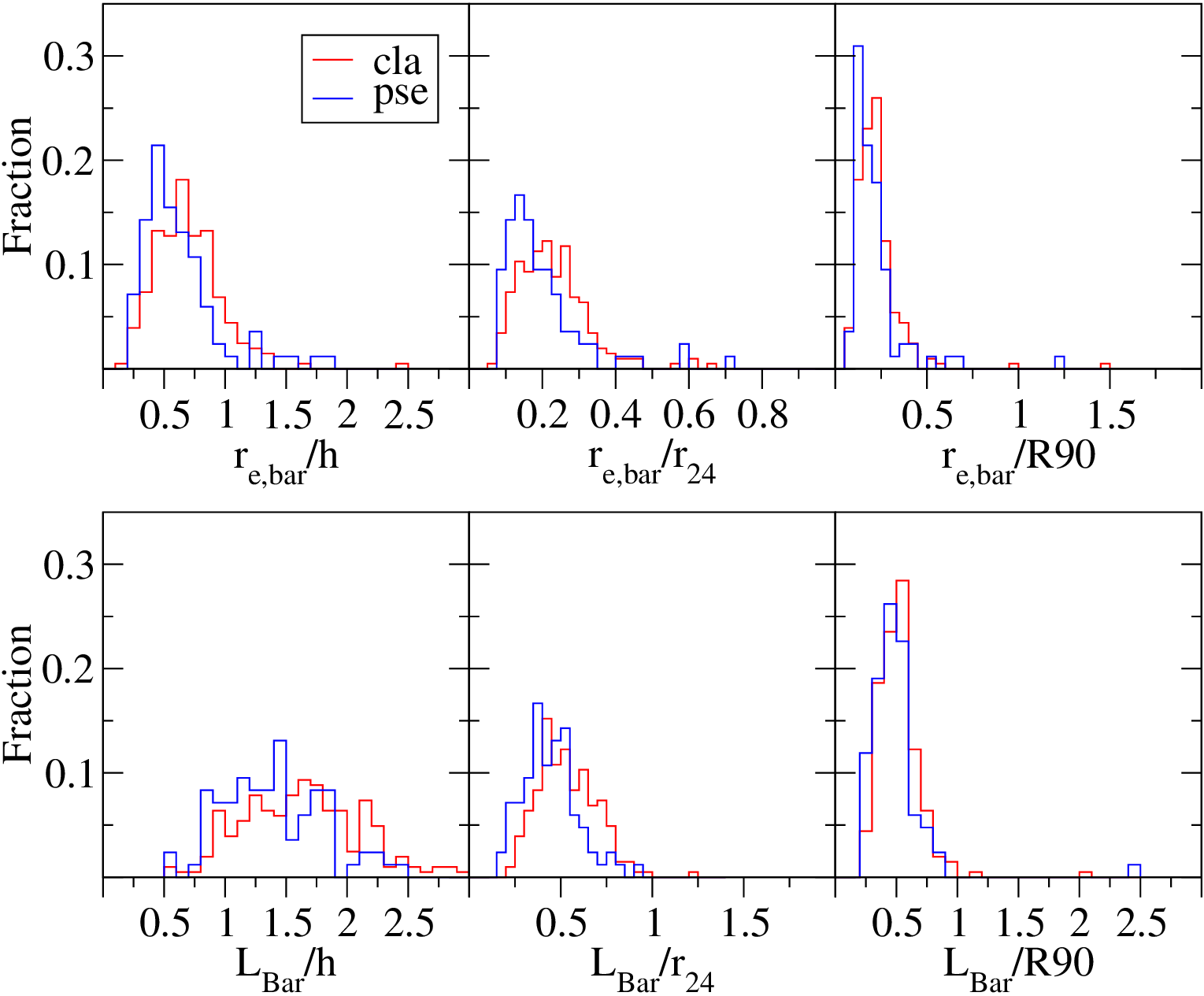}
   \vskip 0.5cm\includegraphics[keepaspectratio=true,width=11cm,clip=true]{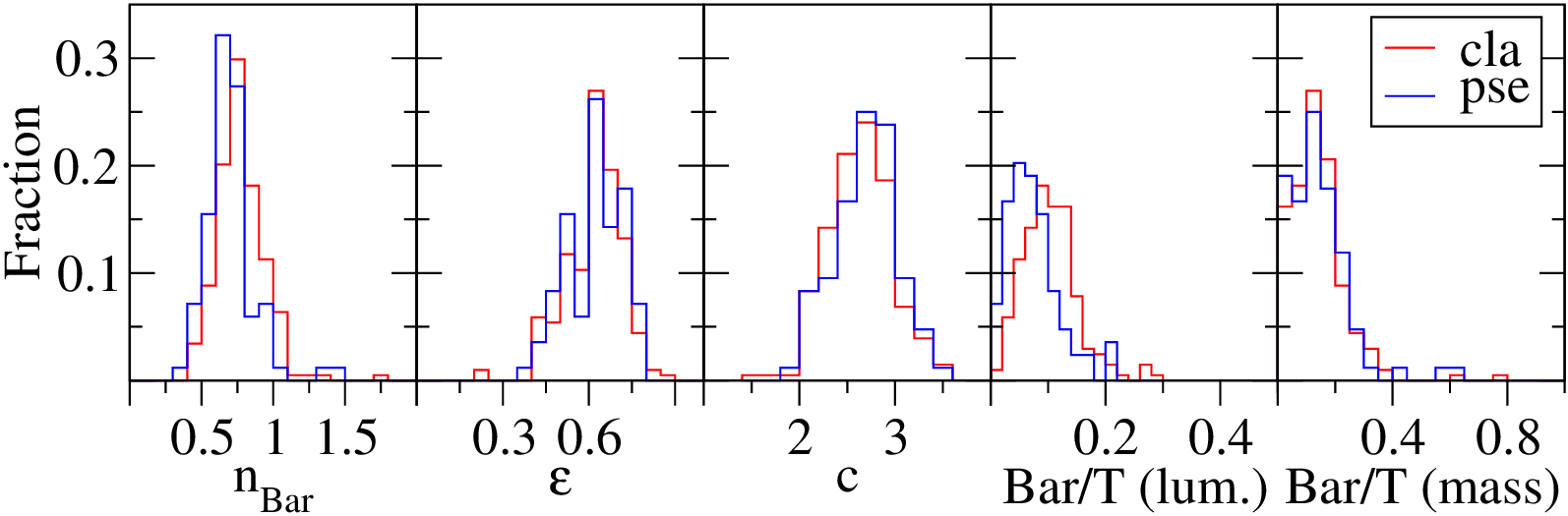}
   \caption{Distributions of bar properties for galaxies with classical and
pseudo-bulges. Bars in galaxies hosting classical bulges tend to be larger than those
in galaxies with pseudo-bulges. This is seen for all parameters used to define
bar size, although the overlap is significant. The bar-to-total luminosity ratio
is also larger in galaxies with classical bulges. Considering all other measured bar
properties, including bar-to-total {\em mass} ratio, galaxies with classical and
pseudo-bulges host similar bars. The arrows in the $L_{\rm bar}$ panel indicate the median values of the corresponding distribution for classical (red arrow) and pseudo-bulges (blue arrow).}
   \label{fig:clapse}
\end{figure*}

As mentioned in the Introduction, classical bulges seem to be formed in relatively fast and violent processes, such as the merging of smaller units, as opposed to pseudo-bulges, which are thought to be formed from disc instabilities, such as bars. It is thus important to ask if bars in galaxies hosting pseudo-bulges are in any sense different from their counterparts in galaxies hosting classical bulges. This question is answered in Fig. \ref{fig:clapse}. The only meaningful difference I find concerns bar length. Bars in galaxies hosting classical bulges tend to be longer than bars in galaxies with pseudo-bulges. This is consistent with previous results that show that bars are longer in early-type disc galaxies \citep[e.g.][]{Erw05,AguMenCor09}, since such galaxies generally host classical bulges, although there are many examples of early-type disc galaxies hosting pseudo-bulges \citep[see e.g.][]{ErwBelGra03,LauSalBut07}. There is also a trend in which bars in galaxies with classical bulges have larger bar-to-total luminosity ratios, but this vanishes when one considers the mass ratio.

\begin{figure}
   \centering
   \includegraphics[keepaspectratio=true,width=8.4cm,clip=true]{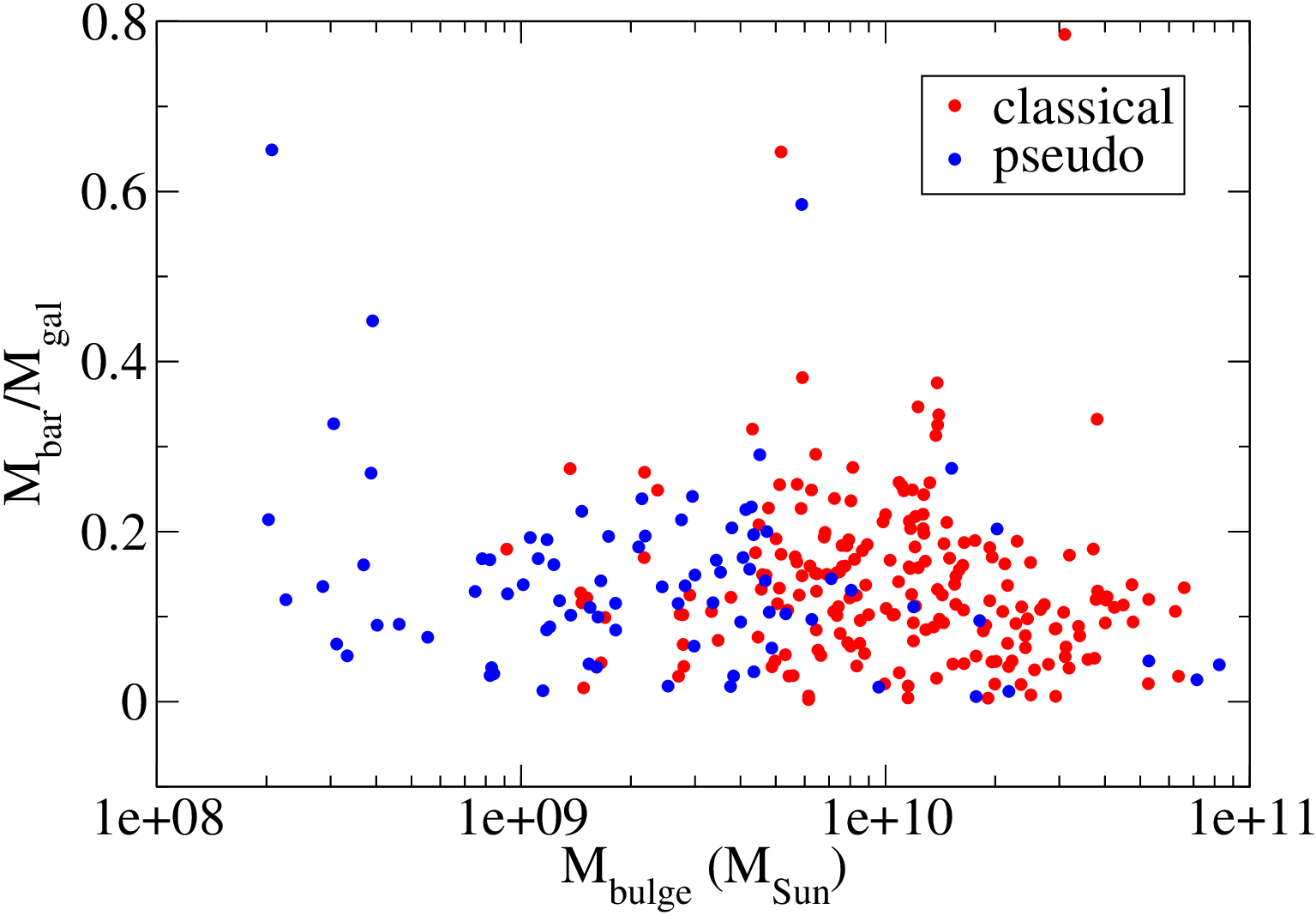}
   \caption{Stellar mass of the bar divided by stellar mass of the galaxy (i.e. the bar-to-total mass ratio) plotted against bulge mass, for classical and pseudo-bulges, as indicated. Clearly, for a fixed bar-to-total mass ratio, pseudo-bulges are on average less massive than classical bulges.}
   \label{fig:mbarmgal}
\end{figure}

One can also ask whether pseudo-bulges relate to their corresponding bars differently than classical bulges, and this is explored in Fig. \ref{fig:mbarmgal}. It shows that, for a fixed bar-to-total mass ratio, pseudo-bulges are on average clearly less massive than classical bulges. Evidently, this relation arises from {\bf (i):} as shown above, most bar properties are unrelated to bulge category, in particular $Bar/T$ (mass), and {\bf (ii):} the separation between bulge categories tend to pick less massive bulges as pseudo-bulges, and, conversely, more massive bulges as classical bulges. Note, however, that pseudo-bulges selected this way also have significantly lower S\'ersic indices and higher star-forming activity, as compared to classical bulges (see Paper I). This indicates that the separation is natural, since such dichotomies are expected between classical and pseudo-bulges. In other words, Fig. \ref{fig:mbarmgal} is just another way of looking at these previous results. It stresses that if pseudo-bulges are formed through disc instabilities such as bars then indeed one needs more than disc instabilities alone to build a classical bulge, as bars in galaxies with classical bulges do not seem more apt to originate more massive bulges than bars in galaxies with pseudo-bulges.

Conversely, if bars are generally quite similar in galaxies with classical and pseudo-bulges, as Fig. \ref{fig:clapse} indeed indicates, then there is no reason to believe that secular effects related to the evolution of the bar do not occur in galaxies with classical bulges. In fact, since bars in the latter tend to be longer, their effects might well be stronger. This suggests that the possibility of having composite bulges, i.e. galaxies hosting both a classical and a pseudo-bulge, is actually quite likely. This has been explored in \citet{Gad09b}, where some bulges were found to have mixed properties: while they appear classical from a structural point of view, the presence of a spectral signature of young stellar populations is typical of pseudo-bulges. Such combination of features naturally appears if one has a more massive and extended classical bulge component, in which a small and young pseudo-bulge is embedded. The classical bulge component produces an imprint in the structural analysis, while the young stellar population from the pseudo-bulge component shows up in the spectral analysis. It seems natural that a disc galaxy can undergo the processes that lead to the formation of a classical bulge {\em and} those that lead to a pseudo-bulge. Curiously, \citet{NowThoErw10} describe two examples of galaxies with composite bulges, in which the classical component is actually smaller.

\subsection{Bar strength and growth}
\label{sec:corr}

In this section, I will look for correlations using the structural properties of bars and their host galaxies that can serve as a test to the theoretical predictions on the secular growth of bars. Essentially, one should check if bars indeed grow longer and stronger in time, and thus one first needs to define bar strength. One readily accessible measure of bar strength is the bar ellipticity. All things equal, a more eccentric bar induces stronger non-axisymmetric forces and torques on the otherwise close to axisymmetric potential of a disc galaxy \citep[see e.g.][and references therein]{Mar95}. Thus, bar ellipticity is a direct measure of bar strength and does not depend on any other galaxy property. A more sophisticated measure of bar strength involves directly estimating the torques induced \citep{BloButKna04,LauSalBut04b,ButVasSal05}. Such torques are normalised by the axisymmetric component of the galaxy potential, which means that, all things being equal, a bar in a galaxy with a more massive bulge is weaker (according to this estimate) than an identical bar in a galaxy with a less massive bulge. The same bar can thus have a different strength, according to this definition, depending on the galaxy it resides. The torque is usually measured at the point in which tangential forces along the bar are at a maximum. Alternatively, it can be measured at other specific points. For instance, \citet{grobutsal10} relate rings and bars using bar forcing at the location of the rings, rather than the position of maximum bar torque. Thus, it can be said that this estimate may be more related to the impact of the bar on the overall galaxy evolution, rather than the strength of the bar itself, and it also depends on which point along the bar the torque is measured, and that can be adapted depending on what is being investigated. Therefore, in the context of this study, it can be argued that bar ellipticity is a more useful, global measure. This is especially true as the theoretical prediction to be tested is the evolution of bar strength along the history of individual galaxies, and the axisymmetric component in the potential of a given (isolated) galaxy is not likely to change in time as much as the difference in such component for different galaxies.

Interestingly, however, bar torques are very well correlated with bar ellipticity \citep{LauSalRau02,BloButKna04}, and therefore the results presented below are not likely to depend significantly on how bar strength is evaluated. It should also be noted that a recent study \citep{ComMarKna09} suggests that a combination of both bar torque and ellipticity can be a useful measure of bar strength. These authors found that the curvature of dust lanes along bars, which is also related to bar strength \citep[see][]{Ath92}, is better related with such combination than with bar torque alone.

\begin{figure}
   \centering
   \includegraphics[keepaspectratio=true,width=6cm,clip=true]{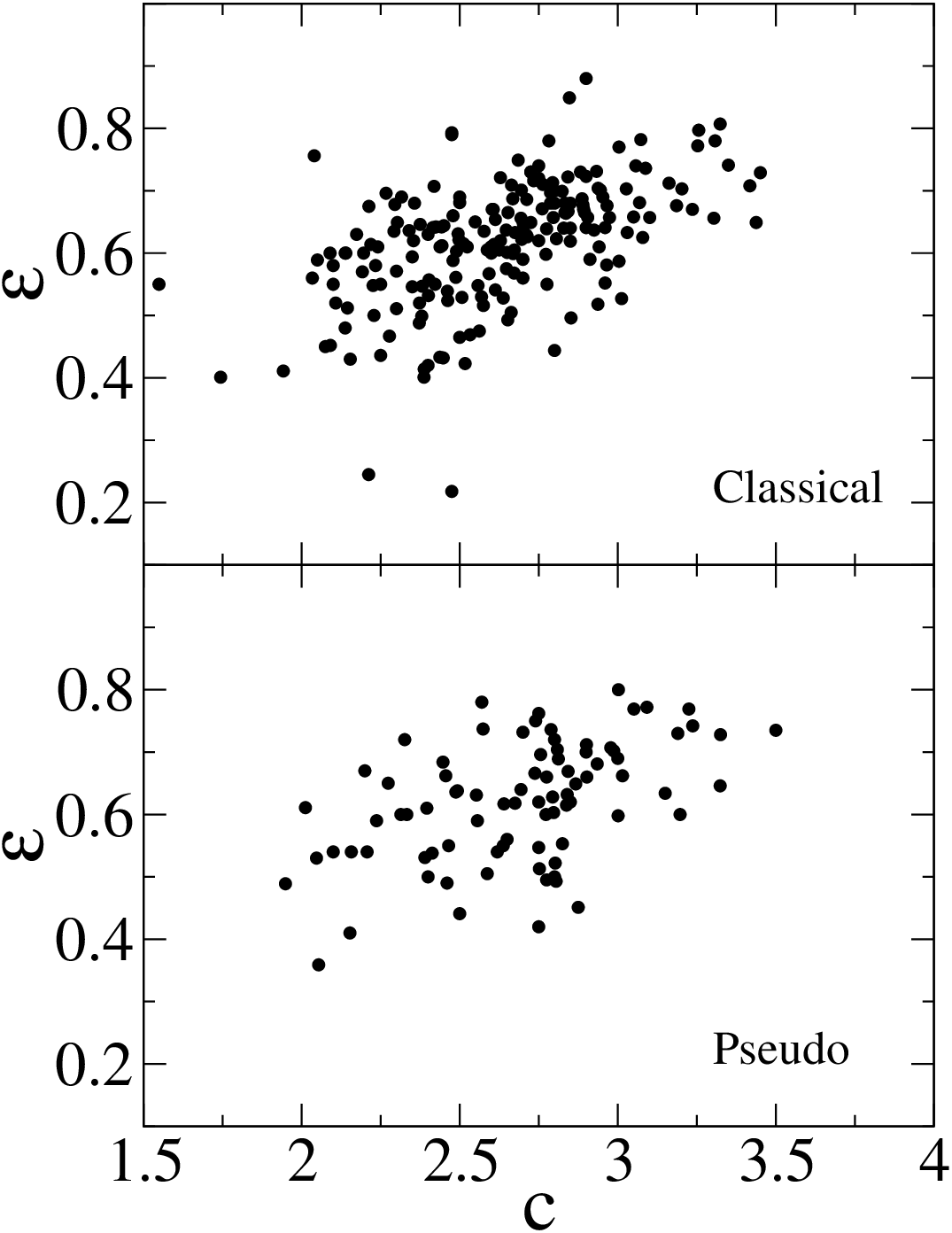}
   \caption{Correlation between bar ellipticity and boxiness for galaxies with
classical and pseudo-bulges, as indicated.}
   \label{fig:ellc}
\end{figure}

Evidently, bar mass, length and even bar boxiness are related to bar strength. In fact, the simulations in \citet{AthMis02} indicate that bars grow stronger in time by getting longer, more eccentric and more boxy, which is consistent with the results from Fig. \ref{fig:ellc}, which shows that bar ellipticity and boxiness are correlated (for both galaxies with classical and pseudo-bulges).
Since both parameters are related to bar strength, I will use the product of the two, i.e., $\epsilon\times c$, as a proxy. It is important to note that the trends discussed below are stronger when one uses $\epsilon\times c$ rather than either $\epsilon$ or $c$ alone. However, there is no improvement if one uses $\epsilon\times c$ times the bar mass or the bar-to-total luminosity and mass ratios. Thus, the choice of using the product $\epsilon\times c$ as a relevant parameter related to bar strength is not arbitrary, but based on the fact the results become clearer by taking the boxiness $c$ into account, as opposed to other parameters. A comparison between $\epsilon\times c$ and bar torque is beyond the scope of the present paper and will be published elsewhere.

\begin{figure*}
   \centering
   \includegraphics[keepaspectratio=true,width=7cm,clip=true]{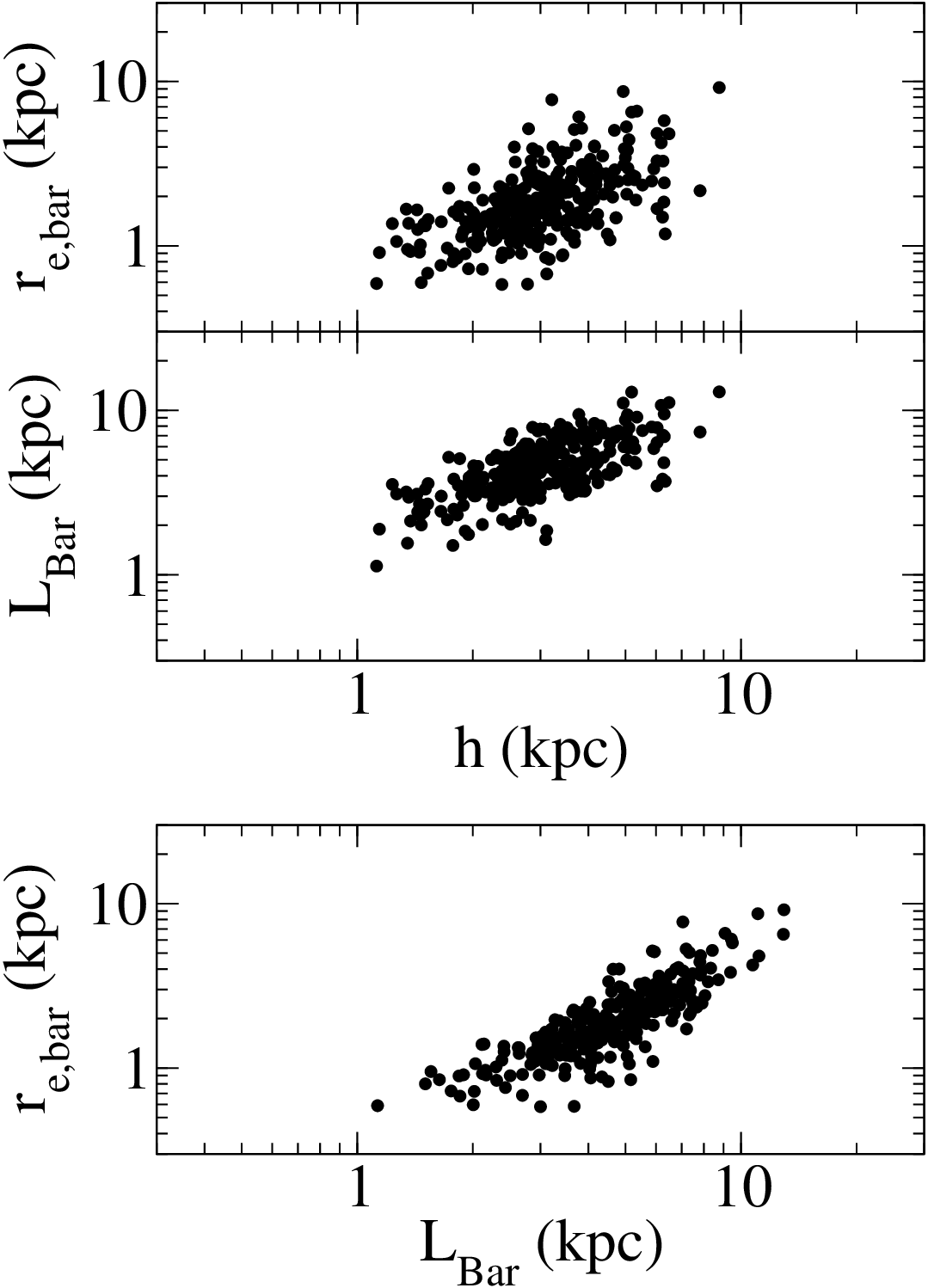}
   \hskip 0.5cm\includegraphics[keepaspectratio=true,width=7cm,clip=true]{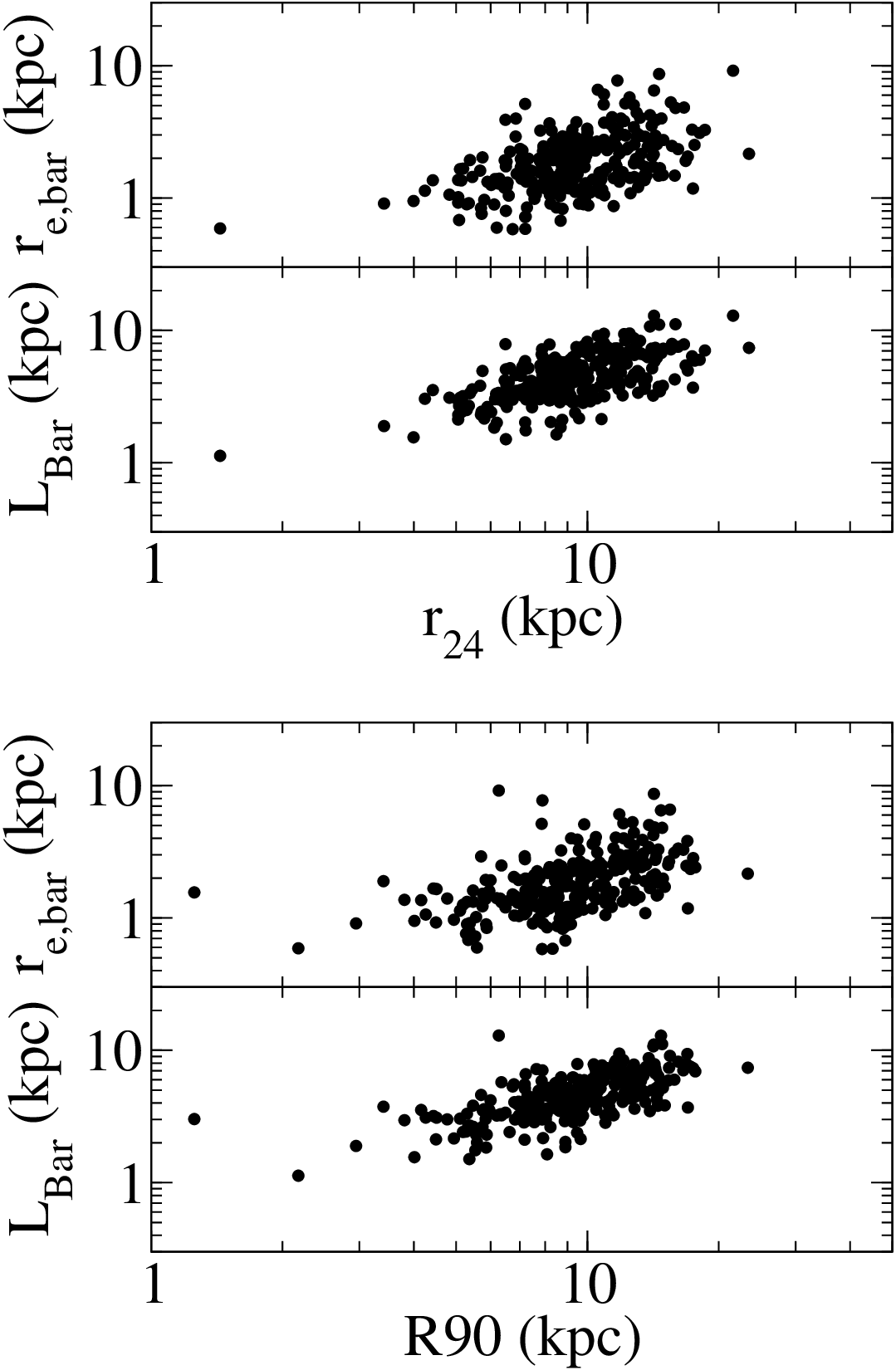}
   \caption{Correlations between the different parameters used to define bar size. They show that,
as expected, larger bars reside in larger galaxies. In addition, bar effective radius is equivalent
to bar length, and $h$, $r_{24}$ and $R90$ can be similarly used to obtain normalised bar sizes.}
   \label{fig:barsizes}
\end{figure*}

I now need a working definition for bar size. Figure \ref{fig:barsizes} shows, consistently with previous results, that bar size correlates with galaxy size, and this is regardless of whether one uses the bar semi-major axis $L_{bar}$, or the bar effective radius $r_{e,bar}$, for the former and $h$, $r_{24}$ or R90 for the latter. It also shows that $L_{bar}$ and $r_{e,bar}$ are well correlated, as expected. Because larger bars are in larger galaxies, a definition for bar size in the context of this study has to be normalised by galaxy size. Figure \ref{fig:barsizes} shows that one can use any combination between $L_{bar}$ or $r_{e,bar}$ and $h$, $r_{24}$ or R90 to define the normalised bar size.

\begin{figure*}
   \centering
   \includegraphics[keepaspectratio=true,width=8.8cm,clip=true]{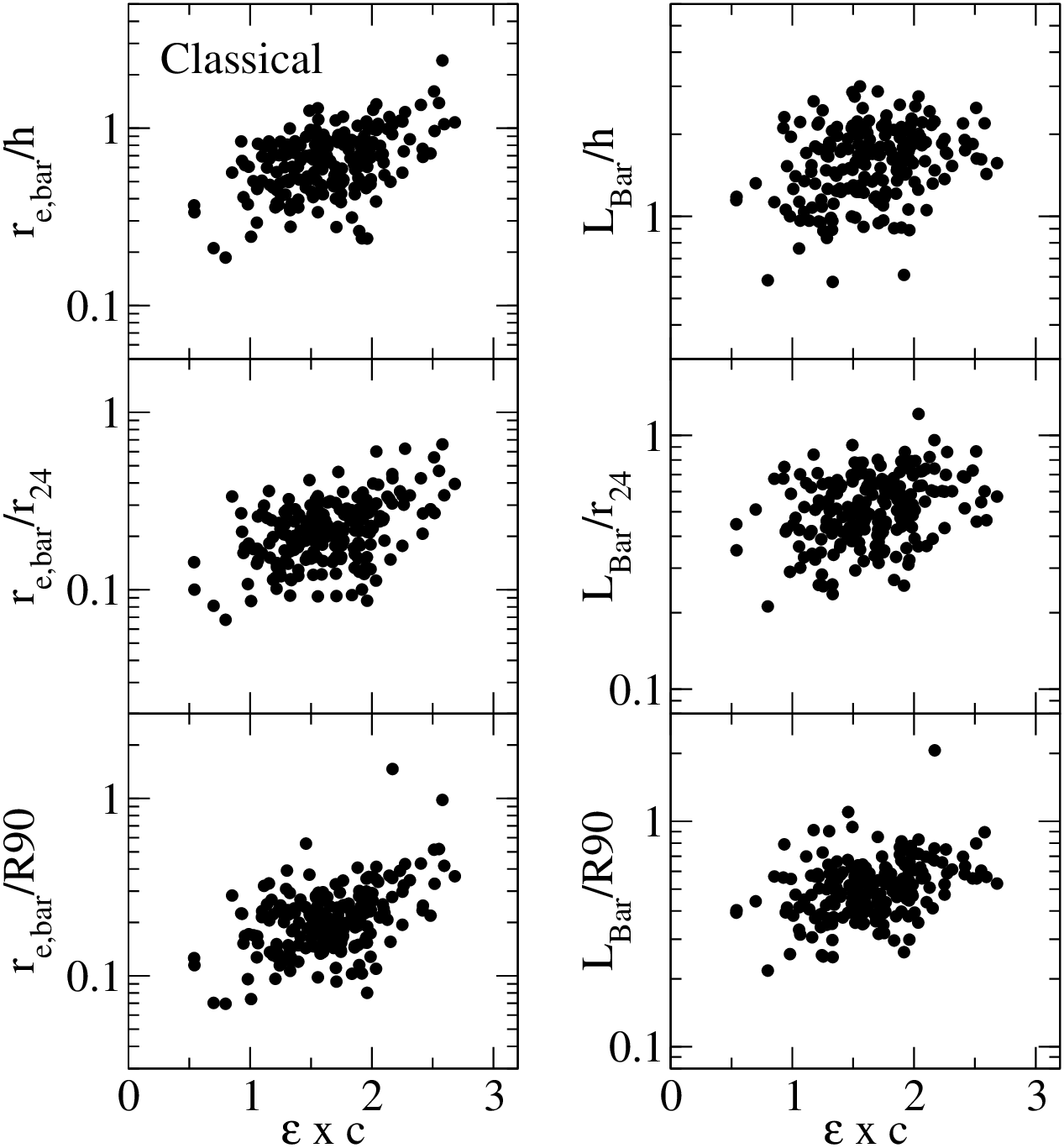}
   \includegraphics[keepaspectratio=true,width=8.8cm,clip=true]{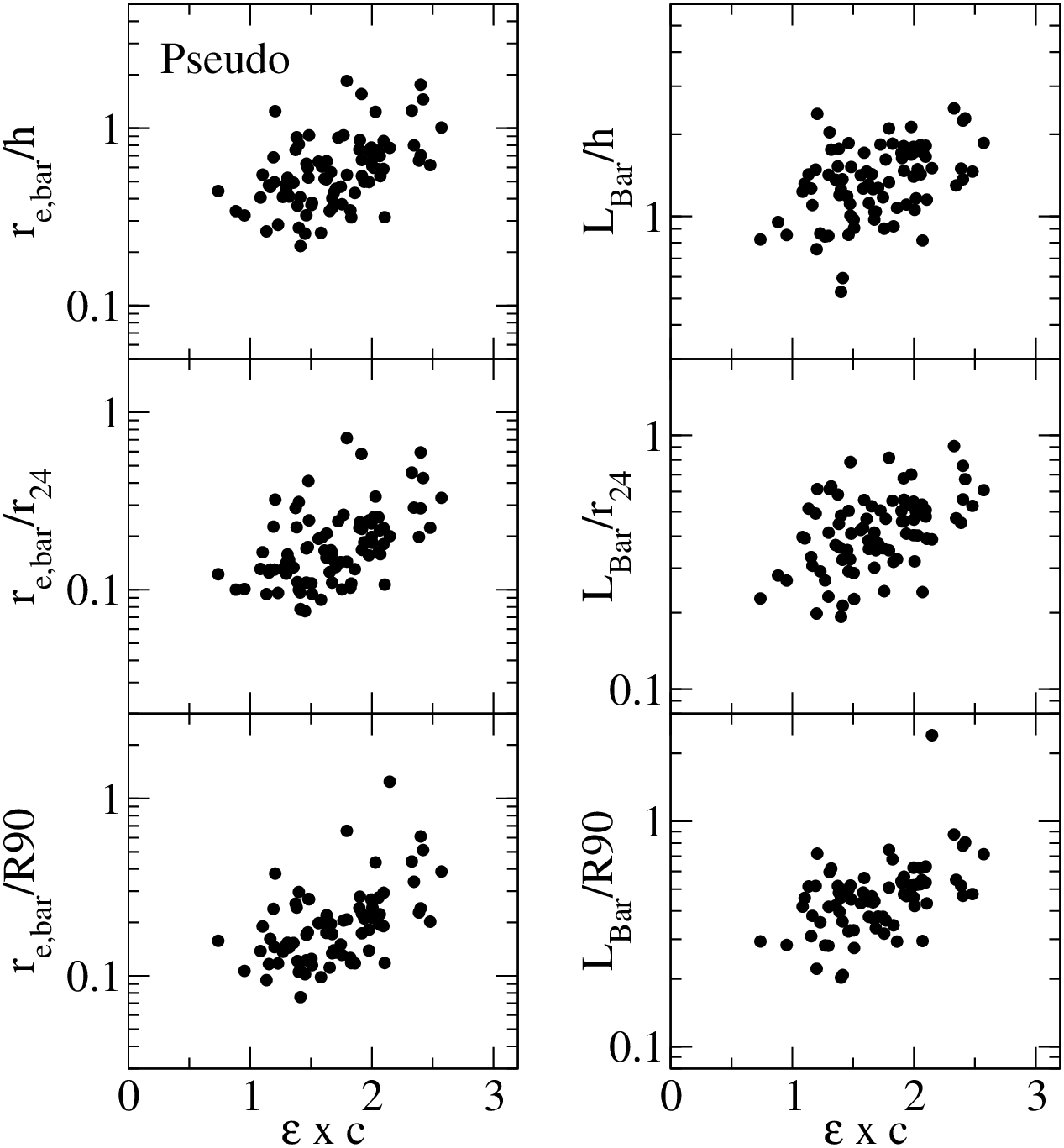}
   \caption{All different measures of normalised bar size plotted against bar $\epsilon\times c$
for galaxies with classical and pseudo-bulges, as indicated. Although a few plots display
large scatter, the positive trend between the parameters is clear.}
   \label{fig:barss}
\end{figure*}

Figure \ref{fig:barss} shows that there is a positive trend between all applicable definitions of normalised bar size and bar $\epsilon\times c$ for both galaxies with classical and pseudo-bulges. This is consistent with the expectations from theoretical work, although clearly there is considerable scatter in some of the plots.

\begin{figure}
   \centering
   \includegraphics[keepaspectratio=true,width=8.4cm,clip=true]{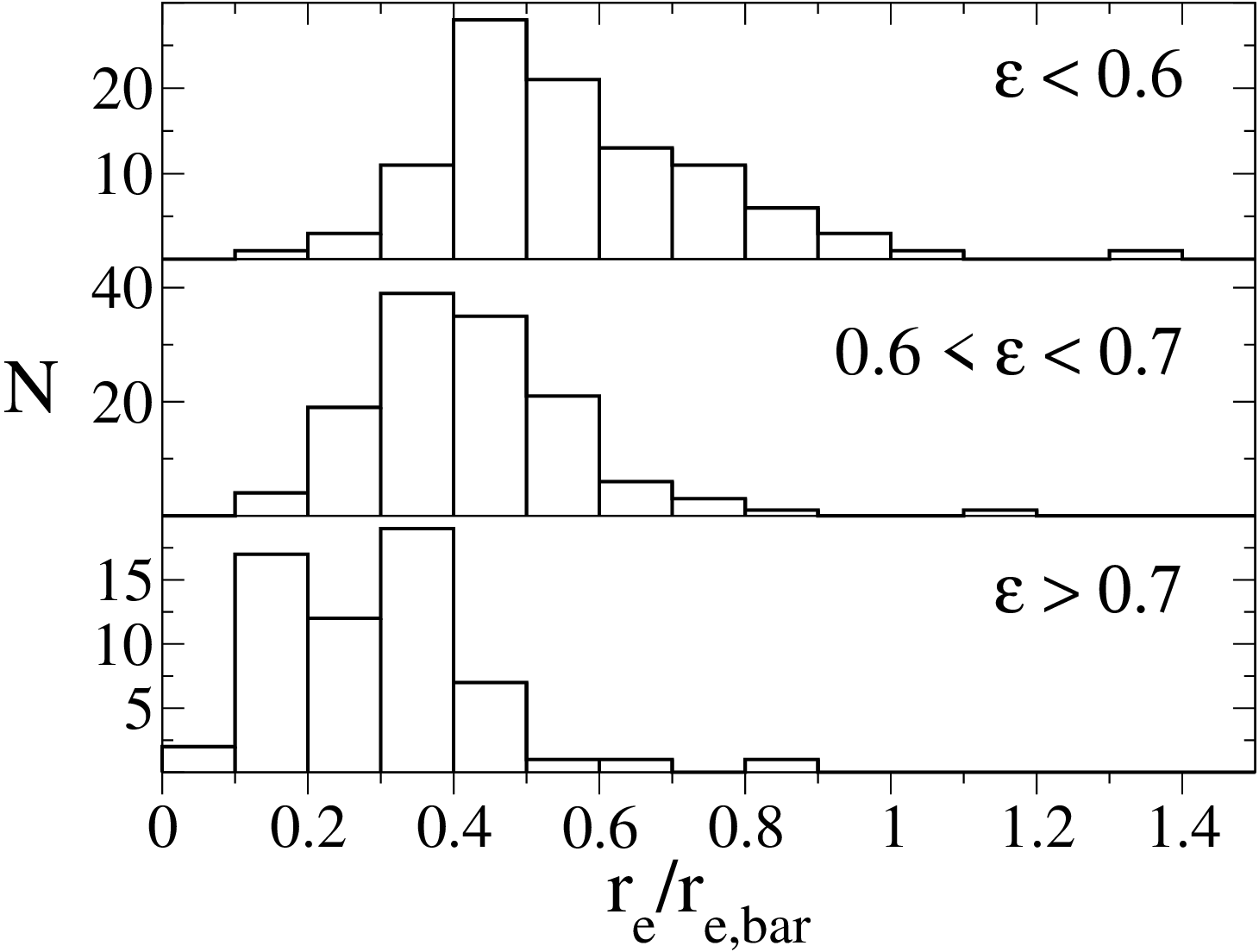}
   \caption{Histograms of the ratio between bulge effective radius and bar effective radius
for bars in three bins of ellipticity, as indicated. It is clear that both parameters are correlated, at
least when $\epsilon<0.7$. Note that the peaks in these distributions shift successively to lower values from bars with $\epsilon<0.6$ to bars with $0.6<\epsilon<0.7$ and finally to bars with $\epsilon>0.7$.}
   \label{fig:rebarbulge}
\end{figure}

Figure \ref{fig:rebarbulge} shows that bar size also correlates with bulge size \citep[see also][]{AthMar80}, which is perhaps not too surprising, but, interestingly, this correlation depends on bar ellipticity. In fact, the correlation is weaker for bars with $\epsilon>0.7$. Surprisingly, $r_e/r_{e,bar}$ shifts to lower values for more eccentric bars. This effect is consistent with the previous finding that more eccentric, stronger bars are longer, and will be further explored below.

We have already seen that longer bars have higher $\epsilon\times c$ values, but the theoretical work I aim to test here also predicts that such bars are dynamically old, i.e. they had more time to evolve. This would happen, for instance, to a bar that resides in a galaxy that have reached a dynamically matured state earlier.\footnote{This assumes that bars are long-lived structures and not recurrent as in the picture proposed by \citet{BouCom02}.} Given the current cosmological picture of galaxy evolution, more massive galaxies reach this point earlier \citep[something referred to as ``downsizing'' -- see discussion in][see also \citealt{camcaroes10}]{SheElmElm08}. But more massive galaxies tend to be those in which the bulge-to-total ratio is larger \citep[see e.g.][]{Gad09b}, and thus if the theoretical work is correct -- and if one assumes that $B/T$ indicates bar dynamical age -- one should find a trend in the sense that longer bars are found in galaxies with larger $B/T$. This is consistent with previous observational results that longer bars tend to be in early-type galaxies, but since I have measurements of $B/T$ I can test it directly.

\begin{figure*}
   \centering
   \includegraphics[keepaspectratio=true,width=8cm,clip=true]{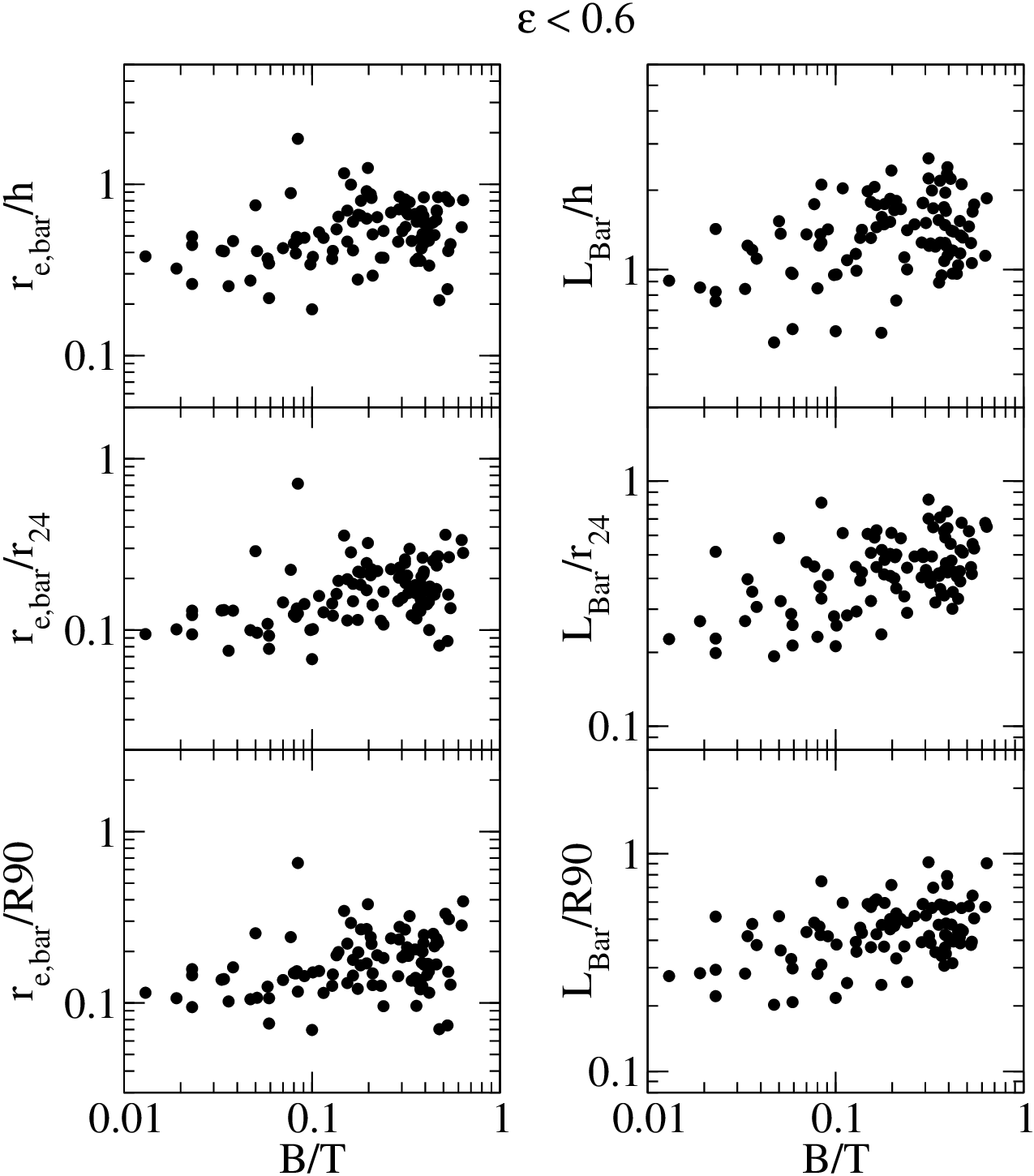}
   \includegraphics[keepaspectratio=true,width=8cm,clip=true]{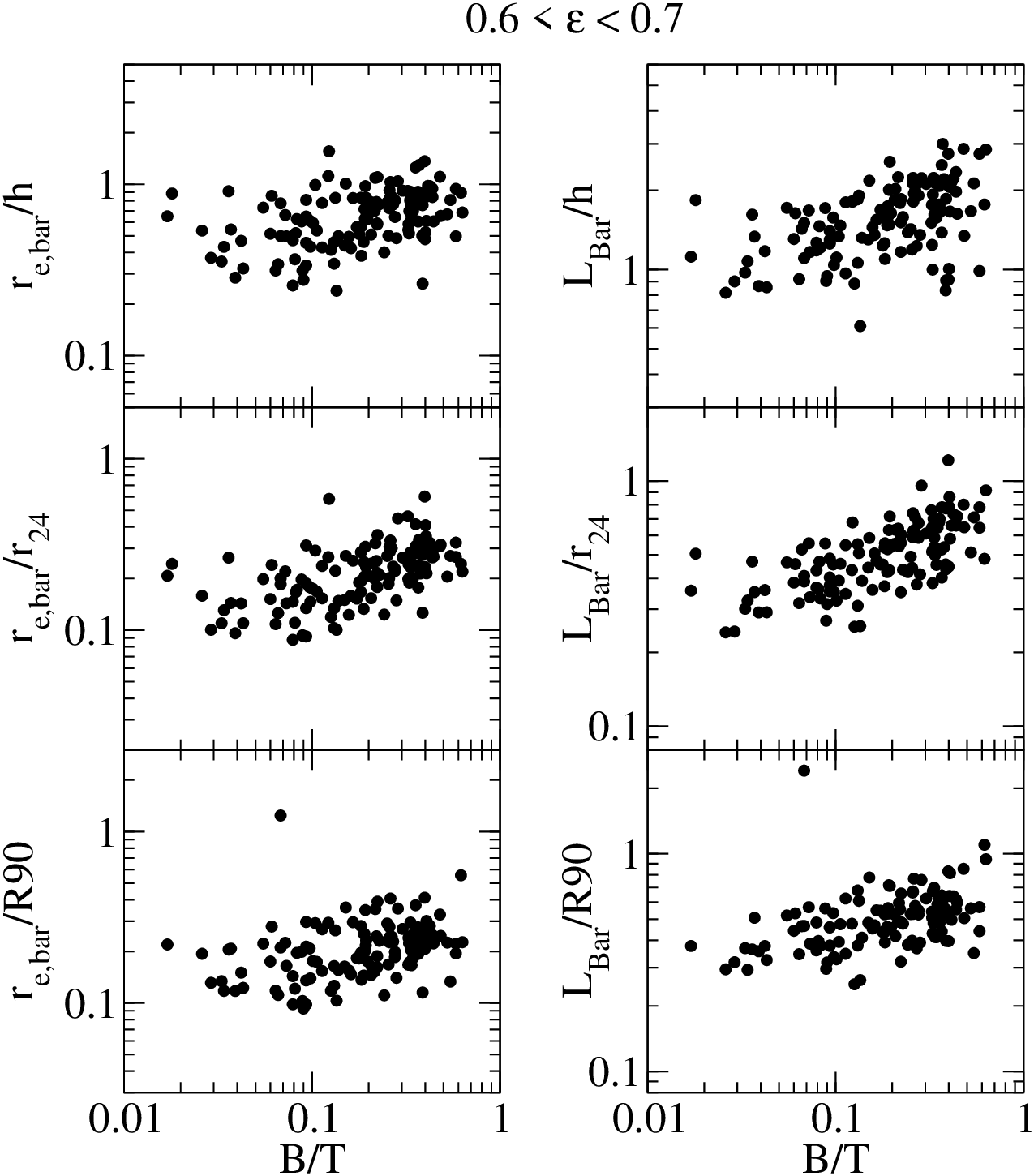}
   \includegraphics[keepaspectratio=true,width=8cm,clip=true]{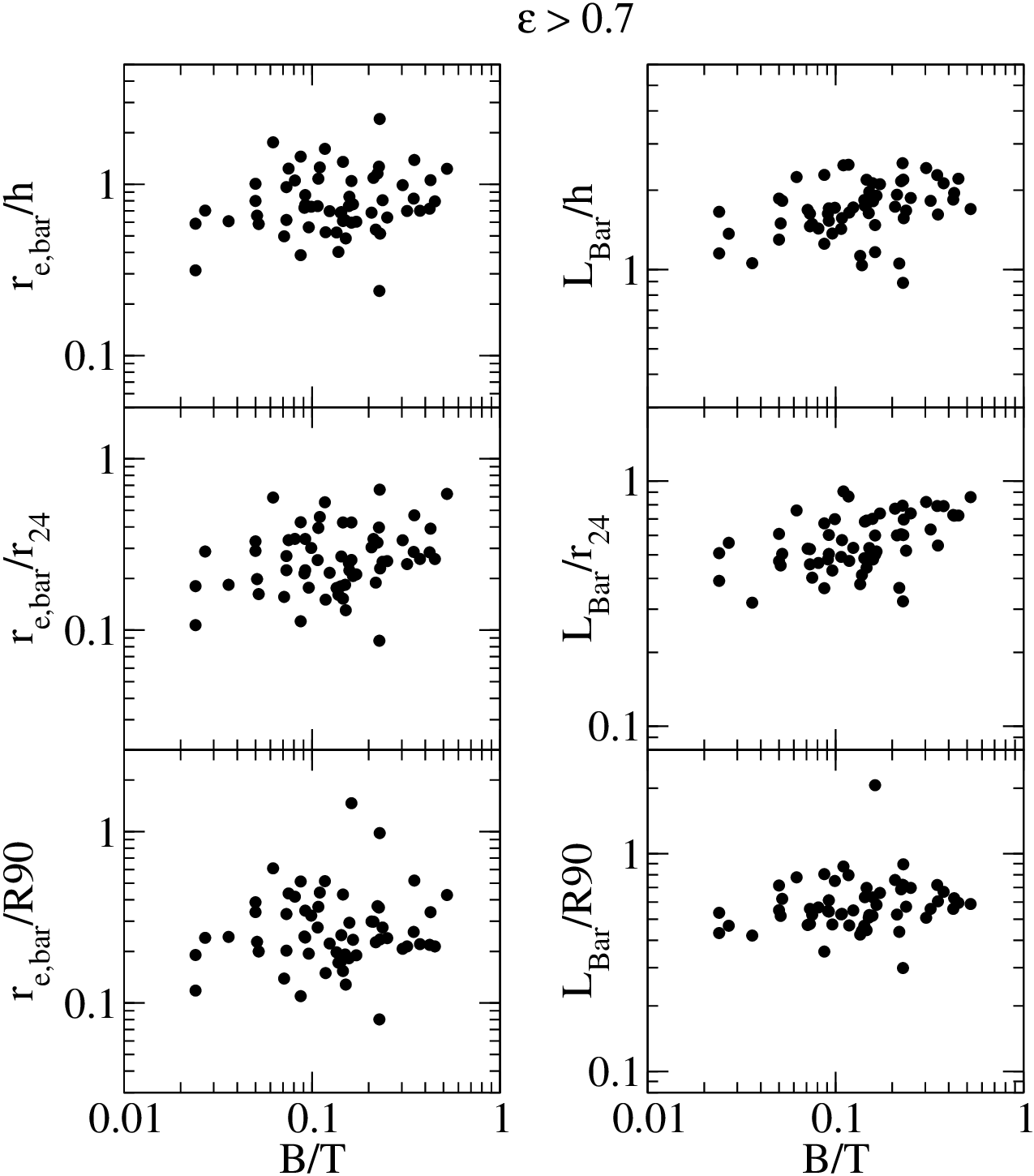}
   \hskip 2cm\includegraphics[keepaspectratio=true,width=5.5cm,clip=true]{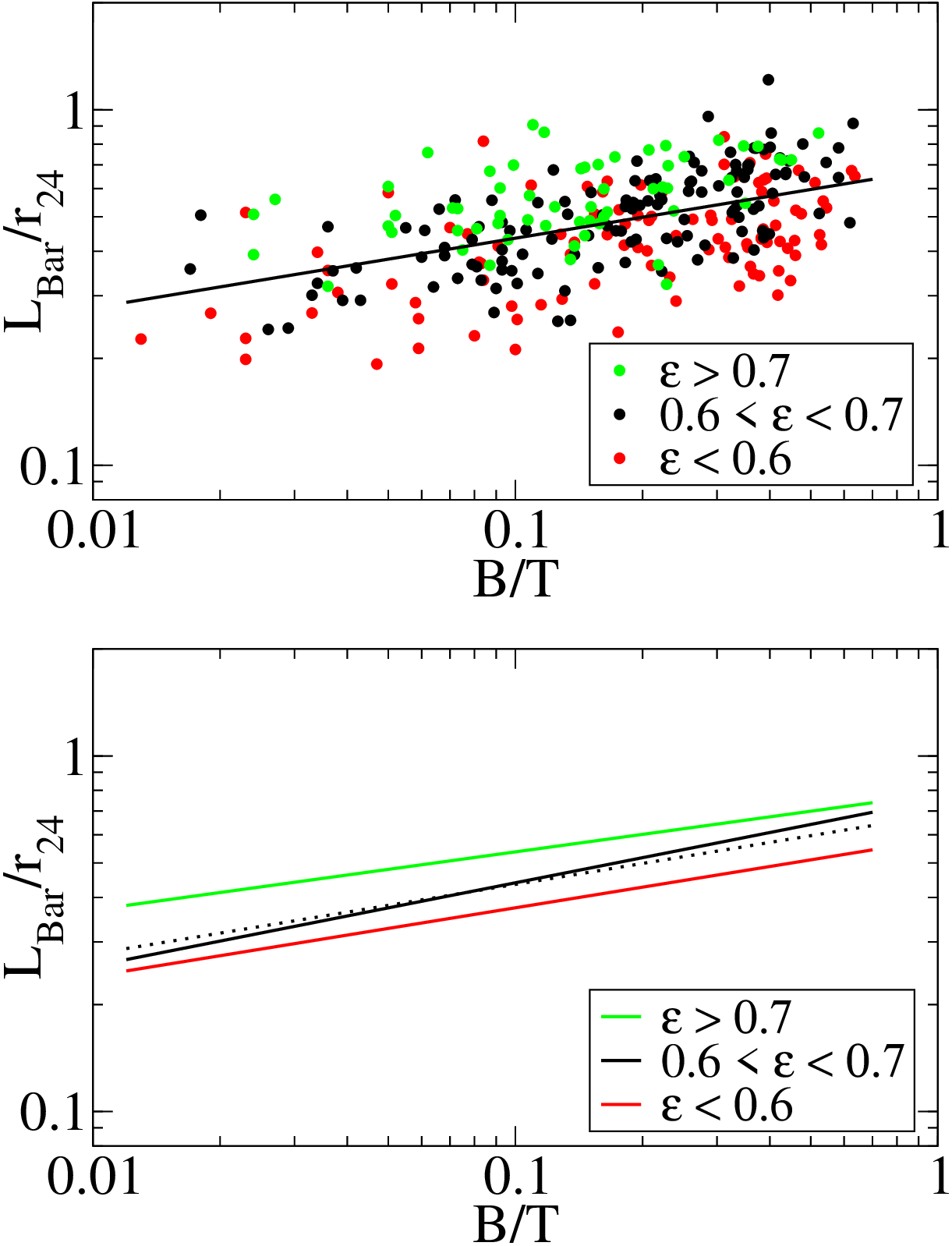}
   \caption{All different measures of normalised bar size plotted against bulge-to-total
ratio, for bars in three bins of ellipticity, as indicated. Although some panels show
considerable spread, in most of them a clear relation is seen in the sense that
longer bars tend to reside in galaxies with more conspicuous bulges. Bars with different
ellipticities describe roughly parallel lines in this relation. This is better seen in the lower
right panels: the top one shows the data for all bars in the panels where $L_{Bar}/r_{24}$ is plotted
against $B/T$, colour coded by bar ellipticity (the solid line is a fit to all points); the
bottom one shows separate fits in the different ellipticity bins (the dotted line in the bottom panel is also a fit to all points, plotted again for reference).}
   \label{fig:barsbt}
\end{figure*}

Figure \ref{fig:barsbt} shows the different measures of normalised bar size plotted against bulge-to-total
ratio. This is done for bars in three bins of ellipticity, as in Fig. \ref{fig:rebarbulge}, for reasons that will be clear shortly below. Although there is considerable scatter in some of these plots, one sees clearly that longer bars tend to be hosted by galaxies with more conspicuous bulges, which is thus consistent with the theoretical expectations. If these plots are done with all bars grouped together, regardless of their ellipticities, this relation is more difficult to see. The reason for that is that bars with different ellipticities follow somewhat parallel lines in these plots, and their combined scatter dilutes the relation. This is easier to see in the bottom right panels in Fig. \ref{fig:barsbt}. The top panel shows $L_{Bar}/r_{24}$ plotted against $B/T$, with colour coding indicating bar ellipticity. The black solid line is a fit to all points. Most green points (bars with $\epsilon>0.7$) are above this line, whereas most red points (bars with $\epsilon<0.6$) are below this line. Black points, representing bars with $0.6<\epsilon<0.7$, lie along the line. The bottom panel shows separate fits to these data points, separated in different ellipticity bins. One sees that the fits describe roughly parallel relations. It should be noted that this effect is apparent for all measures of normalised bar size.

\section{Discussion}
\label{sec:disc}

\subsection{Possible biases}

Some of the physical parameters explored in this study are difficult to measure, such as bar boxiness, in particular considering that the physical spatial resolution of the images used is relatively poor. In \citet{Gad08}, I presented a study using images of 14 nearby galaxies ($z\sim0.005$ - 12 of them barred galaxies) and images of the same galaxies artificially redshifted to $z=0.05$, which results in the same physical spatial resolution of the sample used here. The redshifted and original images were decomposed using the same methodology as in Paper I. By comparing the results from such decompositions, it was verified that parameters such as disc scale length, bulge effective radius, bulge-to-total ratio and bar-to-total ratio can be reliably estimated in the low resolution regime. Bulge parameters are particularly sensitive to seeing effects, though, if $r_e<0.8\times\frac{1}{2}{\rm FWHM}$, but, as shown in Paper I, 97\% of the bulges in the sample have $r_e$ above this threshold.

\begin{figure}
   \centering
   \includegraphics[keepaspectratio=true,width=8.4cm,clip=true]{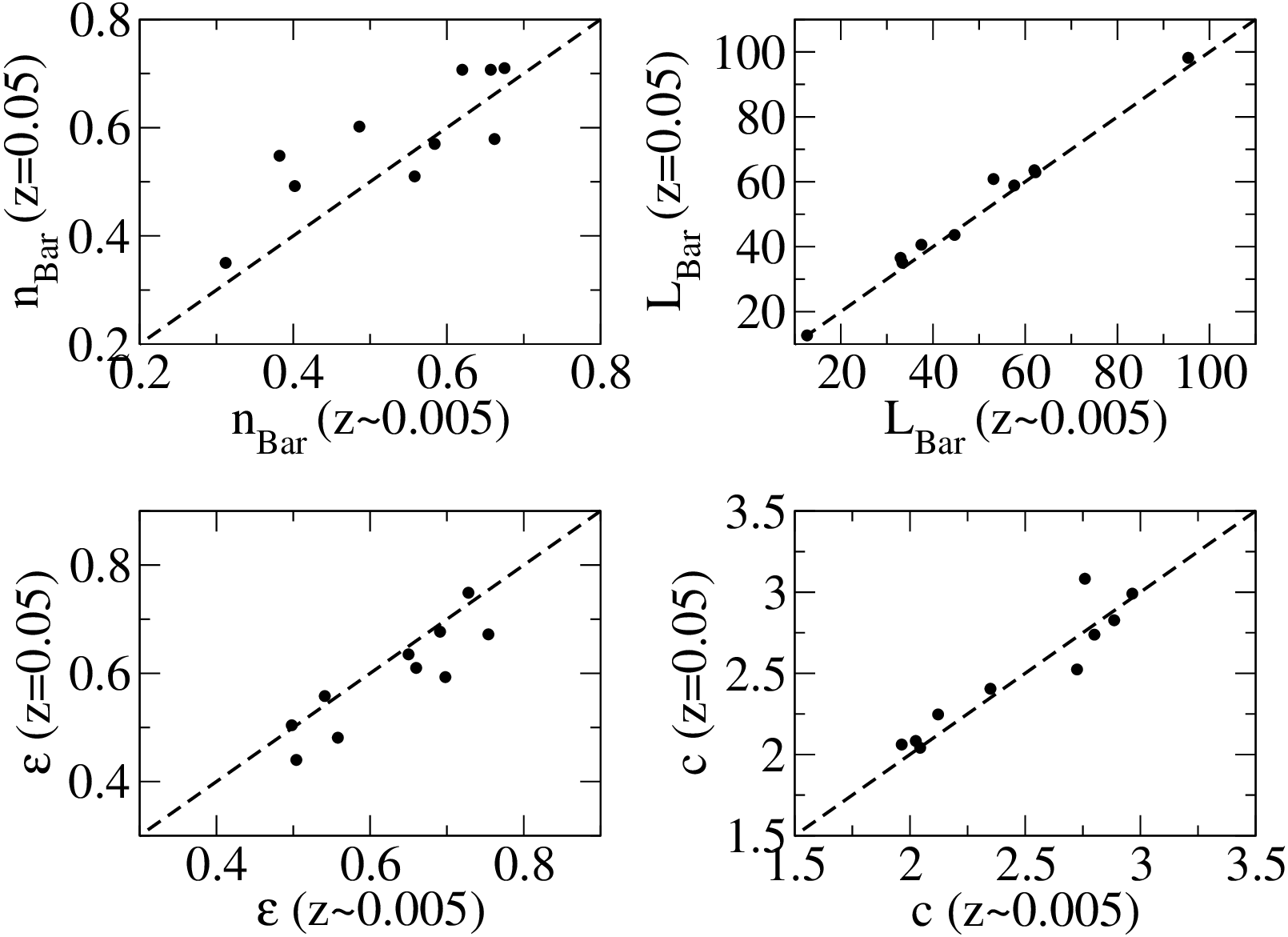}
   \caption{Bar structural parameters, as determined with the redshifted images, plotted against the same parameters obtained with the original images. The dashed lines indicate a perfect correspondence. $L_{\rm Bar}$ is in arcseconds, and measurements from the redshifted images are scaled back to the original galaxy distance. No clear systematic biases are seen, except for a trend in which bar ellipticities, as measured in the redshifted images, have slightly lower values, compared to those measured with the original images. The difference is on average only a few percent, and is the same through the whole range of values obtained, and thus is not harmful to the results presented here.}
   \label{fig:hz_bar}
\end{figure}

I can now use the results from the decompositions in \citet{Gad08} to verify if the other bar structural parameters discussed here can also be reliably obtained in this low resolution regime. Figure \ref{fig:hz_bar} thus compares the estimates for bar S\'ersic index, length, ellipticity and boxiness from the original and artificially redshifted images of the 12 barred galaxies in \citet{Gad08}.\footnote{Note that two bars in this sample become overly faint in the redshifted images and are thus not fitted in these images.} It shows that estimates of such parameters are indeed robust even at the resolution of the images used here. No clear systematic biases are seen, except for a tendency for lower ellipticities, as measured in the redshifted images. This likely results from the rounding and dilution of the bar light at low resolution. It might suggest that the peak in the distribution of bar ellipticity in Fig. \ref{fig:sprops} is slightly offset to lower values. However, the difference is on average only a few percent, and is the same through the whole range of values obtained, and thus not harmful to the results presented here.

One could also be worried that bars might show up more eccentric and boxy in galaxies with less conspicuous bulges, where there is less contamination from bulge light in the bar isophotes. While this is a concern for results based on ellipse fits, it should be stressed that the bulge light is modeled in image decomposition, and thus such concern does not apply here \citep[see fig. 6 and discussion in][]{Gad08}. Further, the results shown above point out the opposite: more eccentric and boxy bars are longer and hosted by bulge-dominated galaxies.

Finally, in the same context, one could argue that higher values of ellipticity and boxiness are easier to obtain in the case of longer bars, since these have a larger fraction of themselves away from the bulge dominant light. However, Fig. \ref{fig:hz_bar} shows that boxiness can be reliably measured at the low resolution regime, i.e. even when bar light is diluted. Furthermore, considering the four galaxies that have their ellipticities underestimated at the low resolution regime (see lower left panel in Fig. \ref{fig:hz_bar}), one sees no tendency for them to be particularly short (these galaxies are NGC 4314, 4394, 4477 and 5701).

\subsection{Bar length in simulations}

As discussed in \citet{Erw05}, to compare bar lengths in observations and simulations one should use $L_{{\rm Bar}}/h$. He provides this value for a number of models from the literature (his Table 8) and argues that there is clearly a lack of short bars in simulations, as compared to observations. Considering only those models in which a live dark matter halo is used \citep{BerHelShl98,AthMis02,ValKly03,HolWeiKat05} one sees that indeed most of them produce bars which are in high end tail of the distribution of $L_{{\rm Bar}}/h$ shown in Fig. \ref{fig:sprops}. In fact, \citet{BerHelShl98} and \citet{AthMis02} report results in which $L_{{\rm Bar}}/h>3$, a value not found in the sample studied here, which is nevertheless biased towards {\em large} bars.

The models that produce the shortest bars are models A$_1$ ($L_{{\rm Bar}}/h=1.4-1.5$), A$_2$ ($L_{{\rm Bar}}/h=1.1-1.3$) and B ($L_{{\rm Bar}}/h=0.8-1.0$) in \citet{ValKly03}. The main difference between models A$_1$ and A$_2$ is that the latter is kinematically colder than the former in the central region. Model B has a smaller halo and a more massive disc. It appears that a less efficient transfer of angular momentum between disc and halo is the cause of these bars being short.

A related issue is that at least in some models \citep[see e.g.][]{MarShlHel06} long bars are formed quickly, i.e. after $1-2$ Gyr only, while the results above suggest that long bars are old. As both bar size and disc size change in time, it would be most useful if theoretical work presents how $L_{{\rm Bar}}/h$ (or $L_{{\rm Bar}}/r_{24}$) evolve with time.

\subsection{Secular evolution of bars}

Although there seems to be some quantitative disagreement between theory and observations concerning bar sizes, a plausible interpretation of the results above suggest a qualitative agreement. The data show that longer bars (using {\em normalised} size) have higher values of $\epsilon\times c$ (Fig. \ref{fig:barss}), and reside in galaxies with higher $B/T$ (Fig. \ref{fig:barsbt}). Now, one can interpret these results, {\em assuming} that higher values of $B/T$ indicate dynamically older bars. This would thus suggest a general trend showing bars getting longer and stronger with time, generally consistent with simulations. This is not the whole picture, though. As discussed in the Introduction, some simulations show that bars can get very large quite fast, quickly become shorter at the onset of the first vertical buckling instability, and then start growing slowly again. Furthermore, simulations taking the effects of gas into account show that the secular growth of bars can be halted, or even reversed, if the gas content is enough.

Since the sample studied here is biased towards large bars, it is plausible that many of these bars have evolved enough, and already went through the first vertical buckling instability. For the same reason, and also because all galaxies in the sample have stellar masses above $10^{10}~{\rm M}_\odot$, the gas content in these barred galaxies is likely on the lower end of the corresponding distribution. This might have helped to make the trends in Figs. \ref{fig:barss}, \ref{fig:rebarbulge} and \ref{fig:barsbt} more apparent. Nevertheless, the complicating effects from gas dynamics and the vertical buckling are probably contributing to the spread seen in these figures. A similar work, with a sample of short bars, in gas-rich system, is likely to shed light on these issues.

With a different methodology, \citet{ElmElmKna07} found that normalised bar size correlates with bar strength and galaxy central density, and also conclude that their results ``suggest that bars grow in length and amplitude over a Hubble time''. 

Observational results like those in e.g. \citet[and references therein]{RauSalLau05} indicate that most bars end near their corotation radius. Along with our current understanding of the orbital structure in barred galaxies, this tells us that bars cannot grow longer if they do not slow down. Therefore, if the interpretation of the results discussed here is correct, namely that bars grow longer with time, one can speculate one step further and suggest that bars also slow down with
time (which is again qualitatively consistent with theory), and with bulge prominence.
Such a relation between the bar pattern
speed 
$\Omega_{\rm B}$ and B/T would be in the sense that bars
rotate slower in galaxies with more prominent
bulges, since these galaxies have longer
bars. This relation could translate to a
dependence of 
$\Omega_{\rm B}$ on Hubble type, although
there is some scatter in the relation between
B/T and Hubble type \citep[e.g.][]{LauSalBut07, GraWor08}.
On direct observational grounds, no solid conclusion can currently be drawn about a dependence of bar pattern speed on Hubble type, and this is due to the difficulty of obtaining reliable estimates of $\Omega_{\rm B}$, in particular for late-type galaxies \citep[see e.g.][]{GerKuiMer03,TreButSal07,Cor10}.

An outstanding and unforeseen new result in this study is the existence of roughly parallel tracks in the correlation between bar normalised size and $B/T$, for bars with different ellipticities (see Fig. \ref{fig:barsbt}). A straightforward way of interpreting
the existence of these parallel tracks is
to conceive that bars could form with different
normalised sizes and ellipticities, and then follow
a somewhat parallel growth. This is a new
aspect that can be investigated with theoretical
work. It would be very interesting to study bars formed with different normalised sizes and ellipticities in simulations, and check whether they follow similar evolutionary paths.

\section{Conclusions}
\label{sec:conc}

I have explored the results from detailed 2D image decomposition of nearly 300 barred galaxies with stellar masses above $10^{10}~{\rm M}_\odot$, at $z\sim0$, concerning mainly the structural properties of bars. This results in a thorough description of bonafide stellar bars in the local universe, including distributions of bar S\'ersic index, ellipticity, boxiness, length and bar-to-total luminosity and mass ratios. The interplay between bars and the bulges and discs in their host galaxies was also examined. Such detailed characterization of local bars can be compared with similar results from studies with samples at higher redshifts, in order to directly investigate how the structure of barred galaxies evolve in time. It can also be compared to results from theoretical work, in order to assess how well theory can describe the structural properties of such bars. Furthermore, a complete description of the properties of local bars is useful to insert ad hoc models of bars in a theoretical framework.

Bars in galaxies hosting classical and pseudo-bulges share similar properties, except that bars in the former are on average larger than those in the latter, considering both absolute and normalised sizes. This is consistent with previous results comparing bar sizes in early- and late-type disc galaxies, as usually the former host classical bulges, whereas the latter host pseudo-bulges. For a fixed bar-to-total mass ratio, pseudo-bulges are on average significantly less massive than classical bulges. This indicates that, if pseudo-bulges are formed through disc instabilities such as bars, then more than that is necessary to build a classical bulge.

Normalised bar size is correlated with the product of the bar ellipticity and boxiness ($\epsilon\times c$, which is related to bar strength) and with bulge-to-total ratio, $B/T$. If higher $B/T$ indicates more evolved bars, then these results can be interpreted as qualitatively consistent with general expectations from theoretical work, which suggests that evolved bars grow longer and stronger in time. This would come along with a decrease in bar pattern speed, and assumes that effects caused by gas dynamics are sufficiently small. Bars with different ellipticities follow parallel tracks in the trend found between normalised bar size and $B/T$, suggesting that bars could form with different normalised sizes and ellipticities but still follow similar evolutionary paths.

\section*{Acknowledgments}
I am indebted to Guinevere Kauffmann for her support throughout this work and useful discussions.
I thank Lia Athanassoula, Albert Bosma and Peter Erwin for important suggestions, and Inma Martinez-Valpuesta for helpful discussions. Remarks from an anonymous referee were very useful to improve the paper. I would also like to thank Enrico Maria Corsini, Victor Debattista and the organisers of the workshop ``Tumbling, Twisting, and Winding Galaxies: Pattern Speeds along the Hubble Sequence'', held in Padova in August 2008, which stimulated intriguing conversations on the subject of this paper.
DAG was supported by the Deutsche Forschungsgemeinschaft priority program 1177 (``Witnesses of Cosmic
History: Formation and evolution of galaxies, black holes and their environment''), and the Max Planck
Society.

The Sloan Digital Sky Survey (SDSS) is a joint project
of The University of Chicago, Fermilab, the Institute for Advanced
Study, the Japan Participation Group, The Johns Hopkins University,
Los Alamos National Laboratory, the Max-Planck Institute
for Astronomy (MPIA), the Max-Planck Institute for Astrophysics
(MPA), New Mexico State University, Princeton University, the
United States Naval Observatory, and the University of Washington.
Apache Point Observatory, site of the SDSS telescopes, is operated
by the Astrophysical Research Consortium (ARC). Funding
for the project has been provided by the Alfred P. Sloan Foundation,
the SDSS member institutions, the National Aeronautics and Space
Administration, the National Science Foundation, the US Department
of Energy, the Japanese Monbukagakusho, and the Max Planck
Society. The SDSS Web site is http://www.sdss.org/.

\bibliographystyle{mn2e}
\bibliography{../../gadotti_refs}

\label{lastpage}

\end{document}